\documentclass[11pt,a4paper,twoside,fleqn]{article}
\usepackage{color,graphicx,natbib,lscape,here,geometry,setspace,multirow,enumitem}
\usepackage[dvipsnames]{xcolor}
\usepackage{graphicx}
\usepackage{authblk}
\usepackage{silence}
\usepackage{rotating}
\usepackage{multirow}

\usepackage{booktabs}
\usepackage{diagbox}
\usepackage{color}
\usepackage{setspace}
\usepackage{algorithm2e}
\usepackage{enumitem}
\usepackage{tikz}

\definecolor{nblue}{HTML}{000660}
\usepackage[colorlinks=true,urlcolor=nblue,linkcolor=nblue,citecolor=nblue]{hyperref}
\usepackage{bigstrut}

\usepackage{tabularx}
\usepackage{threeparttable}
\usepackage{dcolumn}
\newcolumntype{d}[1]{D{.}{.}{#1}}

\usepackage{array}
\newcolumntype{C}[1]{>{\centering\arraybackslash}p{#1}}

\usepackage{etoolbox} 
\makeatletter
\patchcmd{\BR@backref}{\newblock}{\newblock[}{}{}
\patchcmd{\BR@backref}{\par}{]\par}{}{}
\makeatother

\usepackage{pdflscape}
\usepackage{afterpage}

\usepackage{longtable}
\usepackage{array}

\usepackage{comment}
\usepackage{mathtools}

\usepackage[hang,flushmargin]{footmisc} 

\usepackage[hang]{footmisc}
\usepackage[british]{babel}

\pdfoutput=1
\usepackage{float}
\restylefloat{table}

\usepackage[title,titletoc]{appendix}
\makeatletter

\renewenvironment{appendices}{%
    \begin{oldappendices}%
    \renewcommand{\thefigure}{\ifnum \c@section>\z@ \thesection.\fi\@arabic\c@figure}%
    \@addtoreset{figure}{section}%
    \renewcommand{\thetable}{\ifnum \c@section>\z@ \thesection.\fi\@arabic\c@table}%
    \@addtoreset{table}{section}}{%
    \end{oldappendices}%
}\makeatother

\usepackage{titlesec} 
\titleformat{\section}[block]{\large}{\thesection. }{0em}{\MakeUppercase} 
\titleformat{\subsection}[block]{\large}{\thesubsection. }{0em}{\itshape} 
\titleformat{\subsubsection}[block]{\large}{}{0em}{\itshape} 

\let\natbibcitet\citet
\renewcommand\citet{\bibpunct{(}{)}{,}{a}{,}{,}\natbibcitet}
\let\natbibcitep\citep
\renewcommand\citep{\bibpunct{(}{)}{;}{a}{,}{;}\natbibcitep}
\newcommand{\bi}{\begin{itemize}}
\newcommand{\ei}{\end{itemize}}
\newcommand{\be}{\begin{equation}}
\newcommand{\ee}{\end{equation}}
\defcitealias{ieo14}{IEO, 2014}

\long\def\symbolfootnote[#1]#2{\begingroup%
\def\thefootnote{\fnsymbol{footnote}}\footnote[#1]{#2}\endgroup}

\makeatletter
\def\ubar#1{\underline{\sbox\tw@{$#1$}\dp\tw@\z@\box\tw@}}
\def\obar#1{\overline{\sbox\tw@{$#1$}\dp\tw@\z@\box\tw@}}
\makeatother

\usepackage{bm}
\usepackage{caption}
\usepackage{subcaption}

\makeatletter\let\p@subfigure\thefigure\makeatother

\floatstyle{plaintop}
\restylefloat{table}

\usepackage{fancyhdr} 
\usepackage{cleveref}
\crefname{chapter}{Chapter}{Chapters}
\crefname{section}{Section}{Sections}
\crefname{subsection}{Section}{Sections}
\crefname{subsubsection}{Section}{Sections}
\crefname{figure}{Figure}{Figures}
\crefname{table}{Table}{Tables}
\crefname{equation}{Equation}{Equations}
\crefname{appendix}{Appendix}{Appendices}
\crefname{appendices}{Appendix}{Appendices}

\crefname{appsec}{Appendix}{Appendices}

\usepackage{catoptions}
\makeatletter

\def\Autoref#1{%
  \begingroup
  \edef\reserved@a{\cpttrimspaces{#1}}%
  \ifcsndefTF{r@#1}{%
    \xaftercsname{\expandafter\testreftype\@fourthoffive}
      {r@\reserved@a}.\\{#1}%
  }{%
    \ref{#1}%
  }%
  \endgroup
}
\def\testreftype#1.#2\\#3{%
  \ifcsndefTF{#1autorefname}{%
    \def\reserved@a##1##2\@nil{%
      \uppercase{\def\ref@name{##1}}%
      \csn@edef{#1autorefname}{\ref@name##2}%
      \autoref{#3}%
    }%
    \reserved@a#1\@nil
  }{%
    \autoref{#3}%
  }%
}
\makeatother

\title{\LARGE{\textbf{Forecasts with Bayesian vector autoregressions under real time conditions}}}
\author{\large{
\uppercase{Michael Pfarrhofer}}\thanks{
\noindent Salzburg Centre of European Union Studies, University of Salzburg. \textit{Address}: M\"{o}nchsberg 2a, 5020 Salzburg, Austria. \textit{Email}: \href{mailto:michael.pfarrhofer@sbg.ac.at}{michael.pfarrhofer@sbg.ac.at}. \textit{Phone}: +43 662 8044 3772. The author thanks Christoph Pfarrhofer, Niko Hauzenberger, Paul Hofmarcher, Florian Huber, Karin Klieber and Anna Stelzer for valuable comments and suggestions, and gratefully acknowledges financial support from the Austrian Science Fund (FWF, grant no. ZK 35).}
\\\vspace*{-0.5em}
\textit{University of Salzburg}}
\date{}


\def\equationautorefname~#1\null{%
  Eq.~(#1)\null
}
\def\equationautorefname~#1\null{
Eq.~(#1)\null
}

\usepackage[utf8, latin1]{inputenc}                 

\begin{document}
\maketitle\thispagestyle{empty}\normalsize\vspace*{-2em}\small

\begin{center}
\begin{minipage}{0.8\textwidth}
\noindent\small This paper investigates the sensitivity of forecast performance measures to taking a real time versus pseudo out-of-sample perspective. We use monthly vintages for the United States (US) and the Euro Area (EA) and estimate a set of vector autoregressive (VAR) models of different sizes with constant and time-varying parameters (TVPs) and stochastic volatility (SV). Our results suggest differences in the relative ordering of model performance for point and density forecasts depending on whether real time data or truncated final vintages in pseudo out-of-sample simulations are used for evaluating forecasts. No clearly superior specification for the US or the EA across variable types and forecast horizons can be identified, although larger models featuring TVPs appear to be affected the least by missing values and data revisions. We identify substantial differences in performance metrics with respect to whether forecasts are produced for the US or the EA.
\end{minipage}
\end{center}

\onehalfspacing\normalsize\renewcommand{\thepage}{\arabic{page}}
\newpage

\section{Introduction}\label{sec:introduction}
Forecasting economic and financial series is of crucial interest to the private sector and for policy makers in central banks and governments. Many contributions in specialized academic journals thus develop new econometric methods, aiming to improve forecast accuracy of existing approaches. The proposed models are benchmarked against established frameworks, and evaluated in terms of their performance for point and density forecasts mostly by means of so called pseudo out-of-sample exercises. 

In these simulations, one splits available data at a specific point in time into an estimation and holdout period, produces forecasts for the relevant horizon based on the estimation period and assesses these forecasts with realized values in the holdout sample. Subsequently, an additional observation is added to the information set, and the procedure is iterated until the end of data in the holdout is reached. Results for vast sets of models can subsequently be compared, and a ``winner'' in terms of point and density forecasts can be chosen on the grounds of various applicable statistics \citep[see, for instance,][]{geweke2010comparing}.

A key difference for forecasting under real time conditions is, however, that data are frequently subject to revisions and that observations for many series are not yet available to the forecaster until the present date upon first release of the data. This implies that practitioners must rely on incomplete information sets, the ``vintages'' published at a specific point in time. By contrast, pseudo out-of-sample forecasts are produced and evaluated using truncated series from a single vintage. Handling real time data and accounting for revisions and missing values is a burdensome task. Is it necessary to address such concerns, or do pseudo out-of-sample exercises suffice to establish the relative performance ordering of different modeling approaches?

In this paper, we assess the sensitivity of forecasts to taking a real time perspective, and seek to identify models that are most robust to data revisions, and situations where data for the most recent periods are not yet available \citep[for related studies, see][]{diron2008short,giannone2008nowcasting,molodtsova2008taylor,SCHUMACHER2008386,BANBURA2011333,banbura2013now,ghysels2014forecasting,krippner2018real}.\footnote{For pioneering work in the context of real time analysis of data, see \citet{diebold1991forecasting} and \citet{croushore2001real}. A comprehensive survey of the related literature is given by \citet{croushore2006forecasting,croushore2011frontiers}.} Rather than focusing on developing methods for predictable measurement errors in initial releases of statistical agencies \citep[see][]{aruoba2008data,STROHSAL2020} as in \citet{kishor2012var} or \citet{cogley2015price}, our interest centers on assessing the robustness of popular modeling approaches that do not explicitly account for data revisions. We ask whether forecast metrics derived from pseudo out-of-sample and real time studies establish the same ordering in terms of forecast performance, and thus agree on model selection.

We investigate datasets for the United States (US) and the Euro Area (EA). The Federal Reserve Bank of St. Louis maintains monthly vintages for the US, where the dataset for a given month only contains observations until the previous month and data are revised frequently \citep[Federal Reserve Economic Data, FRED-MD, see][]{doi:10.1080/07350015.2015.1086655}. A comparable database is available for the EA, with even longer lags in the publication of available series \citep[Euro Area Real Time Database, EA-RTD, see][]{giannone2012area}.

Using a set of popular models reflecting recent trends in macroeconomic forecasting, that is, Bayesian time-varying parameter (TVP) vector autoregressive (VAR) models with stochastic volatility (SV) of different sizes equipped with a flexible shrinkage prior, we investigate relative differences in forecast performance relying on real time information and pseudo out-of-sample exercises. To gauge the role of large information sets while keeping computation feasible, we also use variants of these models augmented by principal components extracted from the high-dimensional datasets that are available \citep[see also][]{bernanke2003monetary}. Moreover, we employ methods for in-sample imputation of missing values.

Our results suggest three main conclusions. \textit{First}, we observe differences in the relative ordering of model performance for point and density forecasts depending on whether real time data or truncated final vintages in pseudo out-of-sample simulations are used for evaluating forecasts. This finding suggests that when developing econometric frameworks for forecasting, special attention has to be paid to the robustness of the proposed models to missing observations and data revisions. Providing methods for this case is crucial for practitioners, since initial data releases are typically incomplete and imperfect. \textit{Second}, depending on the release schedule of the data, we detect differences in the severity of this problem between the US and the EA. Missing values and data revisions play a much more prominent role for the case of the EA. \textit{Finally}, and perhaps unsurprisingly, producing accurate forecasts is substantially harder when relying on real time data. Depending on the variables and forecast horizon of interest, no clearly superior model specification arises. Relatedly, we cannot identify a unique specification that performs best for the US and EA economy. Differences in forecast performance between real time and pseudo out-of-sample designs are typically smaller for larger and more complex models featuring TVPs. 

The rest of the paper is structured as follows. Section \ref{sec:econometrics} briefly presents the econometric framework. Section \ref{sec:data} introduces the datasets and discusses imputation methods for missing data alongside details on model specification. The main findings are discussed in Section \ref{sec:results}. Section \ref{sec:conclusions} concludes.

\section{Econometric framework}\label{sec:econometrics}
Recent approaches in macroeconomic forecasting usually involve high-dimensional multivariate models to extract information from many available series to forecast key variables of interest. Examples are variants of factor or large Bayesian VAR models \citep[see][]{forni2000generalized,stock2002macroeconomic,banbura2010large,giannone2015prior,doi:10.1080/07350015.2016.1256217,carriero2019large}. In addition, forecasters rely on methods to account for nonlinear dynamics and structural breaks. Key features identified to achieve gains in forecast performance, especially for predictive densities, are SVs, and to a lesser degree, TVPs \citep[see, for instance,][]{d2013macroeconomic,KOOP2013185,carriero2016common,doi:10.1002/jae.2555,feldkircher2017sophisticated,huber2020Bayesian}. For the purposes of this paper, we consider TVP-VARs with SV, equipped with a flexible shrinkage prior as in \citet{feldkircher2017sophisticated} and \citet{HKO}.\footnote{To incorporate high-dimensional information while keeping the computational burden such models entail feasible, we also consider variants augmented by principal components. This amounts to extracting principal components from the full information set other than the series treated as observed (depending on the model size) and including them in the vector $\bm{y}_t$.} 

The baseline TVP-VAR-SV specification for an $M$-dimensional vector of endogenous variables $\bm{y}_t$ for $t=1,\hdots,T$ is
\begin{equation*}
\bm{y}_t = \sum_{p=1}^P\bm{A}_{pt} \bm{y}_{t-p} + \bm{\epsilon}_t, \quad \bm{\epsilon}_{t}\sim\mathcal{N}(\bm{0}_M,\bm{\Omega}_t),
\end{equation*}
with $M\times M$-matrices of time-varying regression coefficients $\bm{A}_{pt}$ (for lags $p=1,\hdots,P$), and a Gaussian error term $\bm{\epsilon}_t$ with zero mean and time-varying $M\times M$-covariance matrix $\bm{\Omega}_t$. In the empirical application, we also include an intercept term that we omit here for brevity.

The covariance matrix can be decomposed as $\bm{\Omega}_t = \bm{H}_t \bm{\Sigma}_t \bm{H}_t'$, with $\bm{H}_t$ denoting a lower unitriangular matrix, and $\bm{\Sigma}_t = \text{diag}(\sigma_{1t}, \hdots, \sigma_{Mt})$. The model can be recast in triangular regression form using $\bm{A}_t = (\bm{A}_{1t},\hdots,\bm{A}_{Pt})$ and $\bm{x}_t = (\bm{y}_{t-1}',\hdots,\bm{y}_{t-P}')'$,
\begin{equation}
\bm{y}_t = \bm{A}_t\bm{x}_t + \bm{H}_t\bm{\eta}_t, \quad \bm{\eta}_t\sim\mathcal{N}(\bm{0}_M,\bm{\Sigma}_t).
\end{equation}

Reparameterizing the model and augmenting the individual VAR equations with the preceeding residuals allows to treat the covariances as regression coefficients \citep[see also][]{carriero2019large}, and for equation-by-equation estimation. Here, we exploit the fact that $\bm{H}_{t}^{-1}\bm{\epsilon}_t = \bm{\eta}_t$.

Denoting the $i$th row of $\bm{A}_t$ by $\bm{A}_{i\bullet,t}$ and the $j$th element in the $i$th row of $\bm{H}_t$ by $h_{ij,t}$, the $i$th equation for $i>1$ is
\begin{equation*}
y_{it} = \bm{A}_{i\bullet,t}\bm{x}_t - \sum_{j=1}^{i-1} h_{ij,t}^{-1} \epsilon_{jt} + \eta_{it}, \quad \eta_{it}\sim\mathcal{N}(0,\sigma_{it}).
\end{equation*}
Equivalently, using $\bm{\beta}_{it} = \left(\bm{A}_{i\bullet,t},\{h_{ij,t}^{-1}\}_{j=1}^{i-1}\right)'$ of dimension $K_i\times1$ with $K_i=pM+i-1$ and corresponding $\bm{z}_{it}=\left(\bm{x}_t',\{-\epsilon_{jt}\}_{j=1}^{i-1}\right)'$,
\begin{equation*}
y_{it} = \bm{\beta}_{it}'\bm{z}_{it} + \eta_{it}.
\end{equation*}

The states are assumed to follow a random walk process with zero mean Gaussian innovations and a diagonal $K_i\times1$-covariance matrix $\bm{V}_i=\text{diag}\left(\{v_{ij}\}_{j=1}^{K_i}\right)$,
\begin{equation*}
\bm{\beta}_{it} = \bm{\beta}_{it-1} + \bm{\varepsilon}_t, \quad \bm{\varepsilon}_t\sim\mathcal{N}(0,\bm{V}_{i}).
\end{equation*}

Using a non-centered parameterization of the model \citep{FRUHWIRTHSCHNATTER201085} allows to treat the square root of the innovation variances in $\bm{V}_i$ as additional regressors,
\begin{align*}
y_{it} &= \bm{\beta}_{i0}'\bm{z}_{it} + \bm{\tilde{\beta}}_{it}'\sqrt{\bm{V}_i}\bm{z}_{it} + \eta_{it},\\
\bm{\tilde{\beta}}_{it} &= \bm{\tilde{\beta}}_{it-1} + \bm{u}_{it}, \quad \bm{u}_{it}\sim\mathcal{N}(\bm{0},\bm{I}_{K_i}).
\end{align*}
Here, $\sqrt{\bm{V}_i}=\text{diag}\left(\{\sqrt{v_{ij}}\}_{j=1}^{K_i}\right)$, and $\bm{\tilde{\beta}}_{it}$ with $\tilde{\beta}_{ij,t} = \frac{\beta_{ij,t}-\beta_{ij,0}}{\sqrt{v_{ij}}}$. This feature allows for inducing shrinkage not only on the constant part of the parameters, but also on the amount of time variation in the states.\footnote{Setting the $v_{ij}=0$ and accounting for the reduced number of free parameters in the posterior of the shrinkage prior yields estimates for constant parameter specifications.}

We rely on equation-specific global-local shrinkage via the popular horseshoe prior \citep{carvalho2010horseshoe}. Note that many shrinkage priors can be used on these coefficients \citep[see][]{cadonna2019triple,HKO}. Popular choices also include the Bayesian Lasso \citep{doi:10.1198/016214508000000337}, the Normal-Gamma \citep{griffin2010} or the Dirichlet-Laplace prior \citep{doi:10.1080/01621459.2014.960967}. For a comparison of the empirical shrinkage properties of various priors in VAR models, see \citet{cross2020macroeconomic}.

On the log of the time-varying variances, we impose the following independent AR(1) processes and use the prior setup and algorithm proposed in \citet{KASTNER2014408}:
\begin{equation*}
\log\sigma_{it} = \mu_i + \phi_i(\log\sigma_{it-1}-\mu_i) + \varsigma_i\xi_{it}, \quad \xi_{it}\sim\mathcal{N}(0,1).
\end{equation*}
All competing models feature stochastic volatilities, based on inferior forecast performance identified in many studies assuming constant variances \citep[see, for instance,][]{clark2011real}. Appendix \ref{app:A} provides details on prior specification and the estimation algorithm.

\section{Real time models for the US and the EA}\label{sec:data}
\subsection{Data sources}
In this paper we consider two available real time datasets for the US and the EA. Both are publicly available for download at \href{https://research.stlouisfed.org/}{research.stlouisfed.org} (Federal Reserve Economic Data, FRED-MD) and \href{http://sdw.ecb.europa.eu/}{sdw.ecb.europa.eu} (Euro Area Real Time Database, EA-RTD), respectively. Detailed descriptions of the data and a priori transformations to achieve approximate stationarity are provided in Appendix \ref{app:B}.

For the US, the Federal Reserve Bank of St. Louis maintains many macroeconomic and financial series on a monthly frequency starting in 1959:01, with monthly data vintages available from 1999:08. We preselect a set of $99$ series with consistent availablity of all relevant variables starting from 1980:01. This establishes the initial sampling period for the forecast exercise with approximately $240$ vintages. The dataset referred to as FRED-MD is described in detail in \citet{doi:10.1080/07350015.2015.1086655}. The publication schedule of this data implies that the vintage in 2000:01, for instance, contains data up to 1999:12, with the current months' values not yet available. Taking a real time perspective, this necessitates these missing values to be ``nowcasted,'' or imputed, to enable forecasts further into the future.

A similar database for the EA is constructed using information from the \textit{Monthly Bulletin} published by the European Central Bank (ECB). This dataset, EA-RTD, is described in \citet{giannone2012area}. The \textit{Monthly Bulletin} provides the ECB Governing Council with the most recent macroeconomic and financial data available, and thus establishes a historical record of vintages. Since many of the variables of interest were only established after the inception of the Euro in 1999, consistent coverage can be achieved starting from 1999:01. We preselect a set of $94$ relevant variables. The first available vintage was published on January 3, 2001, which, accounting for some months where no new data became available or revisions took place twice, results in approximately $180$ vintages per series. To obtain a reasonably long initial estimation period for the forecast exercise, we only rely on vintages published after 2003:12. 

Due to the mode of publication of the underlying \textit{Monthly Bulletin}, the release schedule of the vintages in the EA is less strictly organized than for the case of the US data. Consequently, individual series may exhibit a different number of missing values, depending on the date of the respective release, implying a ``ragged edge'' of data availability at the end of the sample \citep[see also][]{jarocinski2018inflation}. In selected periods, this is also the case for FRED-MD. For instance, while oil prices may be already available for the full length of the vintage sample, inflation data may not yet have been released for the current month. This calls for a conditional nowcast or data imputation scheme, described in the next subsection.

Both datasets range until 2019:01. We preprocess the vintages (when applicable) to account for seasonality using the methods established by the US Census Bureau \citep[X-13-ARIMA-SEATS, see][]{monsell2007x13,sax2018seasonal}, and obtain data on interest rates from the ECB's financial market database. For pseudo out-of-sample simulations, we rely on truncated samples from the final vintage, while for real time simulations we use data available at the specific months. In both cases, we evaluate the forecasts using the final available vintage.

\subsection{Imputing missing values in real time}\label{sec:nowcasting}
The Bayesian approach to estimation allows for a fully probabilistic treatment of missing endogenous variables. There are two relevant cases to be considered. First, a subset of values may be missing, while other variables are already available. Second, at a specific point in time, values for all series may be missing, which allows for sampling these values similarly to unconditional forecasts.

Consider the case where all series are available for $\bm{y}_t$ at time $t$, but a subset of $q$ series at $t+1$ is missing. Let $\bm{y}_{t+1}^{\ast} = (y_{1,t+1}^{\ast},\hdots,y_{q,t+1}^{\ast},y_{q+1,t+1},\hdots,y_{M,t+1})'$ with asterisks indicating missing values and note that the series can always be reordered to yield such a structure. We partition the vector as $\bm{\tilde{y}}_{t+1}^{\ast} = (y_{1,t+1}^{\ast},\hdots,y_{q,t+1}^{\ast})'$ of dimension $q\times1$ and $\bm{\tilde{y}}_{t+1} = (y_{q+1,t+1},\hdots,y_{M,t+1})'$ of size $(M-q)\times1$. 

The fitted values at time $t+1$ are $\bm{A}_{t+1}\bm{x}_{t+1} = (\bm{\mu}_1',\bm{\mu}_2')'$, with $\bm{\mu}_1$ being a $q\times1$ vector corresponding to the missing values, and $\bm{\mu}_2$ the $(M-q)\times1$ vector related to the available series. Similarly, we partition the covariance matrix $\bm{\Sigma}_{t+1}$, denoting the upper left $q\times q$ block by $\bm{\Sigma}_{11}$, the upper right $q\times(M-q)$ block by $\bm{\Sigma}_{12}$, the bottom left $(M-q)\times q$ block by $\bm{\Sigma}_{21}$, and the bottom right $(M-q)\times (M-q)$ block by $\bm{\Sigma}_{22}$. 

The distribution of the missing values conditional on the realizations, the endogenous variables up to $t$ and all other model parameters, follows from the properties of the multivariate Gaussian distribution. In particular,
\begin{align*}
\bm{\tilde{y}}_{t+1}^{\ast}|\bm{\tilde{y}}_{t+1},\bm{y}_{t},\hdots,\bm{y}_{1},\bullet &\sim \mathcal{N}(\bm{\bar{\mu}},\bm{\bar{\Sigma}}),\\
\bm{\bar{\mu}} &= \bm{\mu}_1 + \bm{\Sigma}_{12}\bm{\Sigma}_{22}^{-1}(\bm{\tilde{y}}_{t+1} - \bm{\mu}_2),\\
\bm{\bar{\Sigma}} &= \bm{\Sigma}_{11} - \bm{\Sigma}_{12} \bm{\Sigma}_{22}^{-1} \bm{\Sigma}_{21}.
\end{align*}
Conditioning on the realized values alters both the mean $\bm{\bar{\mu}}$ and variance $\bm{\bar{\Sigma}}$ of the distribution of the missing values, although the variance does not depend upon the particular value of the realizations. For the case where all values at a specific point in time are missing, the distribution of $\bm{\tilde{y}}_{t+1}^{\ast}$ is
\begin{equation*}
\bm{\tilde{y}}_{t+1}^{\ast}|\bm{y}_{t},\hdots,\bm{y}_{1},\bullet \sim \mathcal{N}(\bm{A}_{t+1}\bm{x}_{t+1},\bm{\Sigma}_{t+1}).
\end{equation*}

Nowcasting the missing values in this way is related to data augmentation techniques, and moreover allows for drawing the model coefficients conditional on the synthetic data in each iteration of the algorithm. Consequently, posterior uncertainty surrounding both the missing data and model parameters is accounted for during estimation.

\subsection{Model specification}
This subsection describes the differently sized information sets used for estimation. We focus on forecasting inflation, a short-term interest rate and the unemployment rate as key variables of interest (for abbreviations, see Appendix \ref{app:B}). We rely on three different model sizes, where the information sets are specified to approximately correspond to similar studies employing VAR models for forecasting, conditional on the availability of the respective series in all vintages:
\begin{itemize}[leftmargin=1.5em,itemsep=0em]
	\item \textit{Small} ($M=3$ variables): The small model employs the focus variables. For the US we thus include {CPIAUCSL} as a monthly inflation indicator, {FEDFUNDS} as the key interest rate and {UNRATE} to measure the unemployment rate; for the EA, we use {C\_OV} for inflation, {EUR3M} as short-term interest rate and {UNETO} for the unemployment rate.
	\item \textit{Medium} ($M=6$ variables): This specification incorporates the \textit{small} information set and adds stock prices, industrial production and long-term interest rates. The variable codes for the additional quantities are {S\&P500}, {INDPRO}, {GS10} (US) and {DJE50}, {XCONS}, and {10Y} (EA).
	\item \textit{Large} ($M=11$ variables): The large VAR features the information set of the {small} and {medium} models and further includes information on oil prices, exchange rates, loans, a money aggregate, and a term spread: {OILPRICEx}, {EXUSUKx}, {BUSLOANS}, {M2SL}, {T10YFFM} for the US; and {OILBR}, {ERC0\_BGR}, {LOANSEC\_U\_NG}, {M2\_V\_NC}, {10Y2Y} in the case of the EA.
\end{itemize}

To circumvent scaling issues arising from the data and to improve stability of the models, we standardize each vintage prior to estimation to have zero mean and unit variance by subtracting the mean and dividing all series by their standard deviation. We denote the sample mean and standard deviation by $\bm{m}_{y}$ and $\bm{s}_{y}$, respectively.

Augmented variants of the baseline specification to use the full available information sets are constructed as follows. Depending on the respective model size, we extract five principal components from the remaining (demeaned and standardized) series and include them in the vector of endogenous variables $\bm{y}_t$. Varying the number of principal components does not significantly alter the forecasting results. Rather than taking a fully Bayesian approach, we rely on this approximation to reduce the computational burden. This implies that the largest model features $M=16$ endogenous variables. We use $P=2$ lags.

\section{Forecasts under real time conditions}\label{sec:results}
One of the main questions this paper raises is whether pseudo out-of-sample forecasts suffice to establish a clear ranking in terms of predictive performance that also applies in the real time context. We assess both moments of the predictive distributions and evaluate point and density forecasts. Details on the employed metrics are provided in Appendix \ref{app:A}.

\subsection{Overall forecast performance}\label{sec:results_overall}
To gauge overall forecast performance of the competing models in real time and pseudo out-of-sample designs, we rely on cumulative joint log predictive scores (LPS) over the full holdout. Higher scores indicate superior performance, and allow for constructing a ranking of the models on a monthly frequency. We obtain these ranks for the real time and pseudo out-of-sample exercises, and study their correlation over time to see whether they agree on the ordering of the models in terms of relative forecast performance. 

\begin{figure}[t]
\begin{subfigure}[b]{\textwidth}
\caption{United States}
\includegraphics[width=\textwidth]{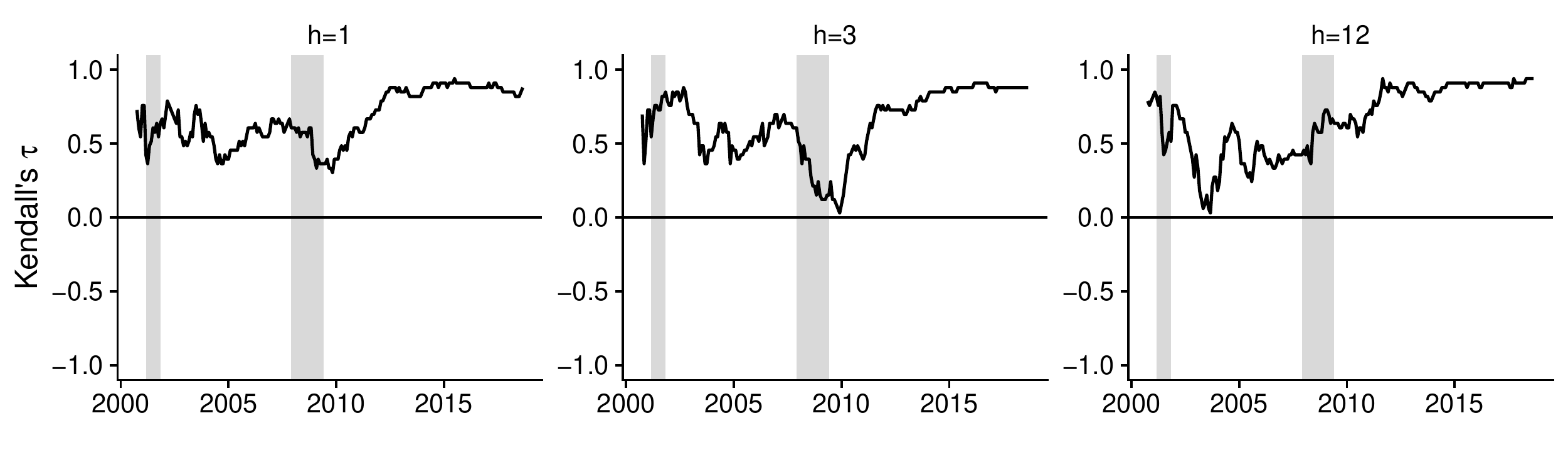}
\end{subfigure}
\begin{subfigure}[b]{\textwidth}
\vspace*{1em}\caption{Euro Area}
\includegraphics[width=\textwidth]{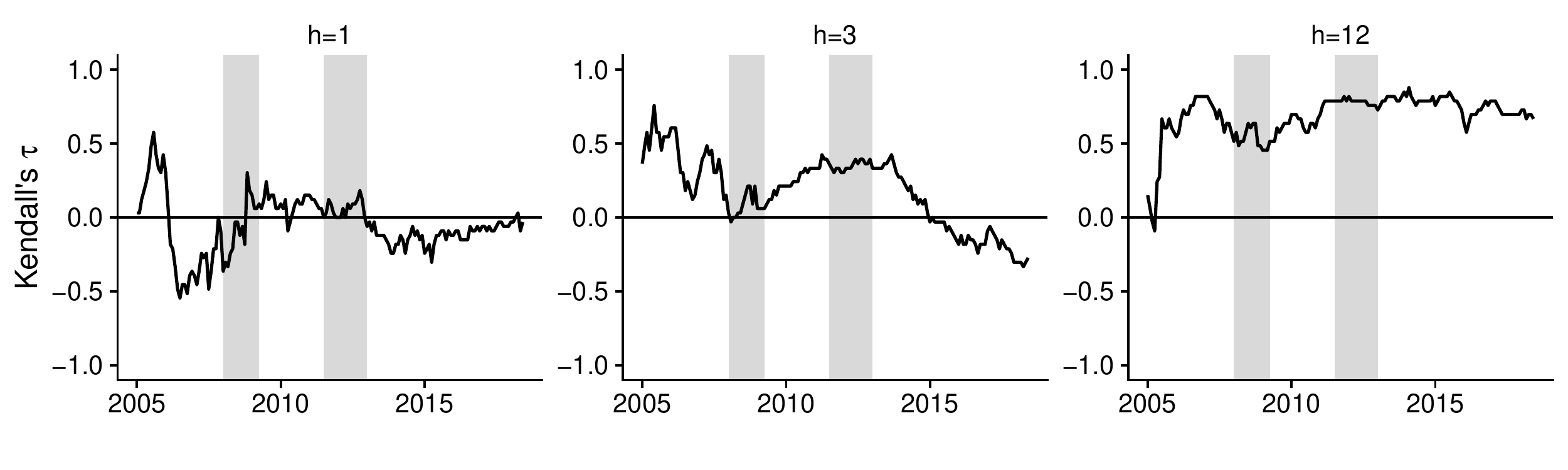}
\end{subfigure}
\caption{Rank correlation of different specifications between real time and pseudo out-of-sample forecasts over the holdout.}\label{fig:jlps_corr}\vspace*{-0.3cm}
\caption*{\footnotesize\textit{Note}: Ranks are derived from cumulative joint log predictive scores. The grey shaded areas indicate recessions dated by the NBER Business Cycle Dating Committee (US) and the CEPR Euro Area Business Cycle Dating Committee (EA).}
\end{figure}

As a summary statistic, we calculate Kendall's rank correlation coefficient $\tau$ for all holdout periods.\footnote{Popular nonparametric correlation estimators are Spearman's $\rho$ and Kendall's $\tau$. We choose Kendall's $\tau$ due to its superior robustness properties \citep{croux2010influence} and note that estimates for the correlation are almost identical in terms of both Spearman's and Kendall's coefficient.} Values of $\tau$ close to one signal that real time and pseudo out-of-sample exercises produce similar rankings, values close to zero indicate no association, while values close to negative one imply reversed orderings. The resulting correlation coefficient over time at the one-month, one-quarter and one-year ahead horizon is depicted in Fig. \ref{fig:jlps_corr} for the US (a) and the EA (b).\footnote{For the US, recessions are dated by the National Bureau of Economic Research (NBER) Business Cycle Dating Committee and published at \href{https://www.nber.org/cycles/cyclesmain.html}{nber.org/cycles/cyclesmain.html}. For the EA, recessions are dated by the Centre for Economic Policy Research (CEPR) Euro Area Business Cycle Dating Committee and published at \href{https://eabcn.org/dc/chronology-euro-area-business-cycles}{eabcn.org/dc/chronology-euro-area-business-cycles}. Since recessions are only dated on a quarterly basis in the EA, we pick the first month of the respective quarter as the reference period.}

Figure \ref{fig:jlps_corr} shows that real time and pseudo out-of-sample information sets produce different rankings in terms of model performance depending on the forecast horizon, and more importantly, specific features of the data releases. Comparing the US and the EA, differences in the release schedule of real time data for the EA lead to substantially more disagreement between real time and pseudo out-of-sample simulations regarding the relative performance of the competing models.

Zooming in on the forecasts for the US, we detect a high concordance of real time and pseudo out-of-sample performance rankings. The rank correlation coefficient is close to one, especially after $2010$. Depending on the forecast horizon, the measure detects subtle differences over time. For one-month ahead forecasts, we observe that the relative ordering seems to change between the two recessions in the holdout, and during the Great Recession. These changes are even more pronounced for the one-quarter ahead forecasts, with Kendall's $\tau$ indicating values close to zero during and just after the Great Recession. Interestingly, for one-year ahead forecasts, differences in the ranking of the models occur mainly between the two recessions.

For the EA, using real time information appreciably changes the relative performance ordering of the competing specifications, especially at short forecast horizons. At the one-month ahead horizon, we observe periods over the full holdout where the model ordering between real time and pseudo out-of-sample information sets is reversed, indicated by negative values of Kendall's $\tau$. During and between the two recessions, the correlation fluctuates around zero. For the one-quarter ahead horizon, the rankings tend to agree with each other early in the holdout. After the European debt crisis, the rank correlation coefficient turns negative, indicating that real time information and pseudo out-of-sample rankings diverge. By contrast, for one-year ahead forecasts, real time and pseudo out-of-sample designs suggest similar rankings of relative performance, with correlations close to one apart from a brief period early in the holdout.

\begin{figure}[!htbp]
\begin{center}
\begin{subfigure}[b]{\textwidth}
\caption{United States}
\includegraphics[width=0.95\textwidth]{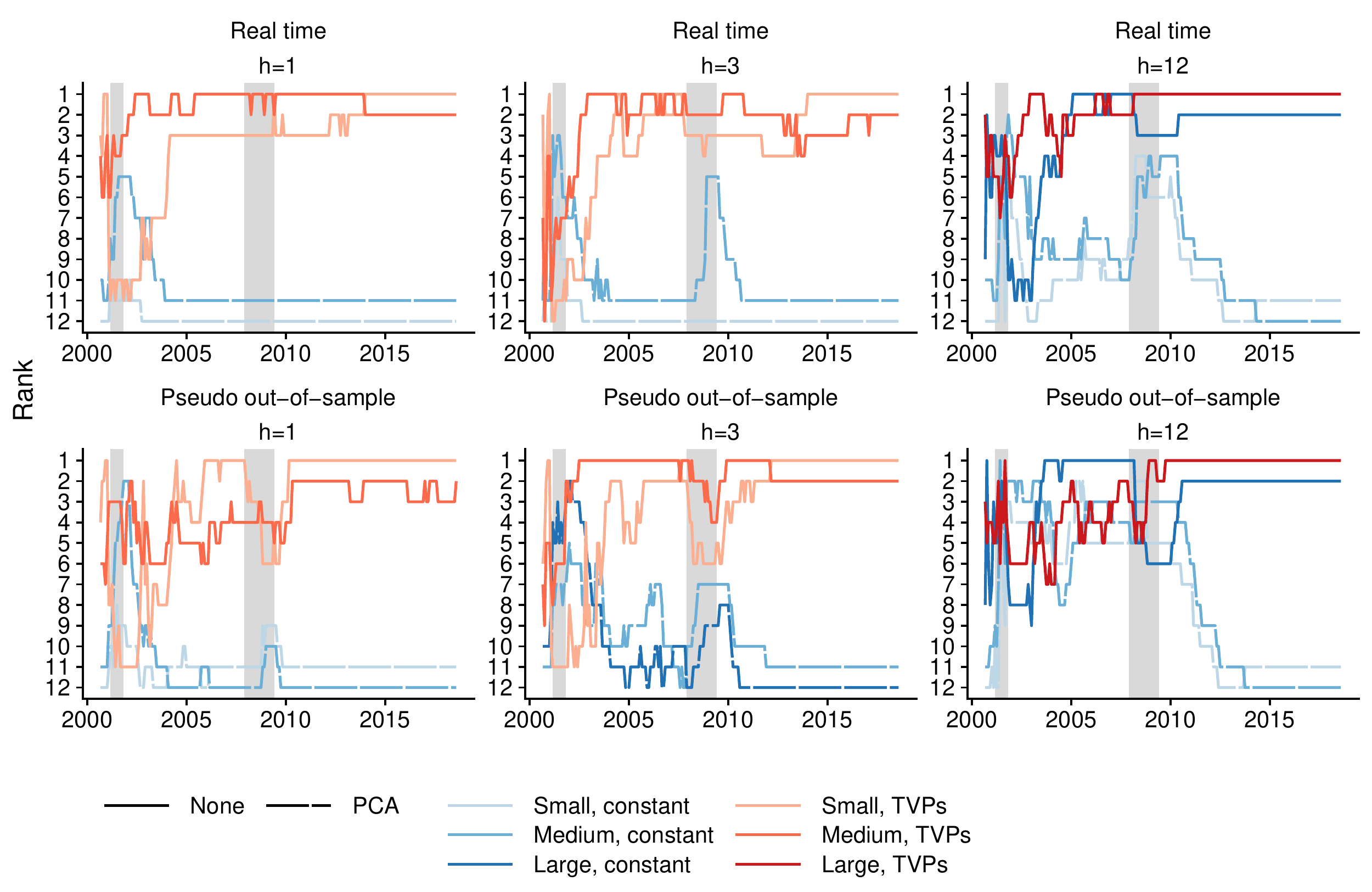}
\end{subfigure}
\begin{subfigure}[b]{0.95\textwidth}
\vspace*{1em}\caption{Euro Area}
\includegraphics[width=\textwidth]{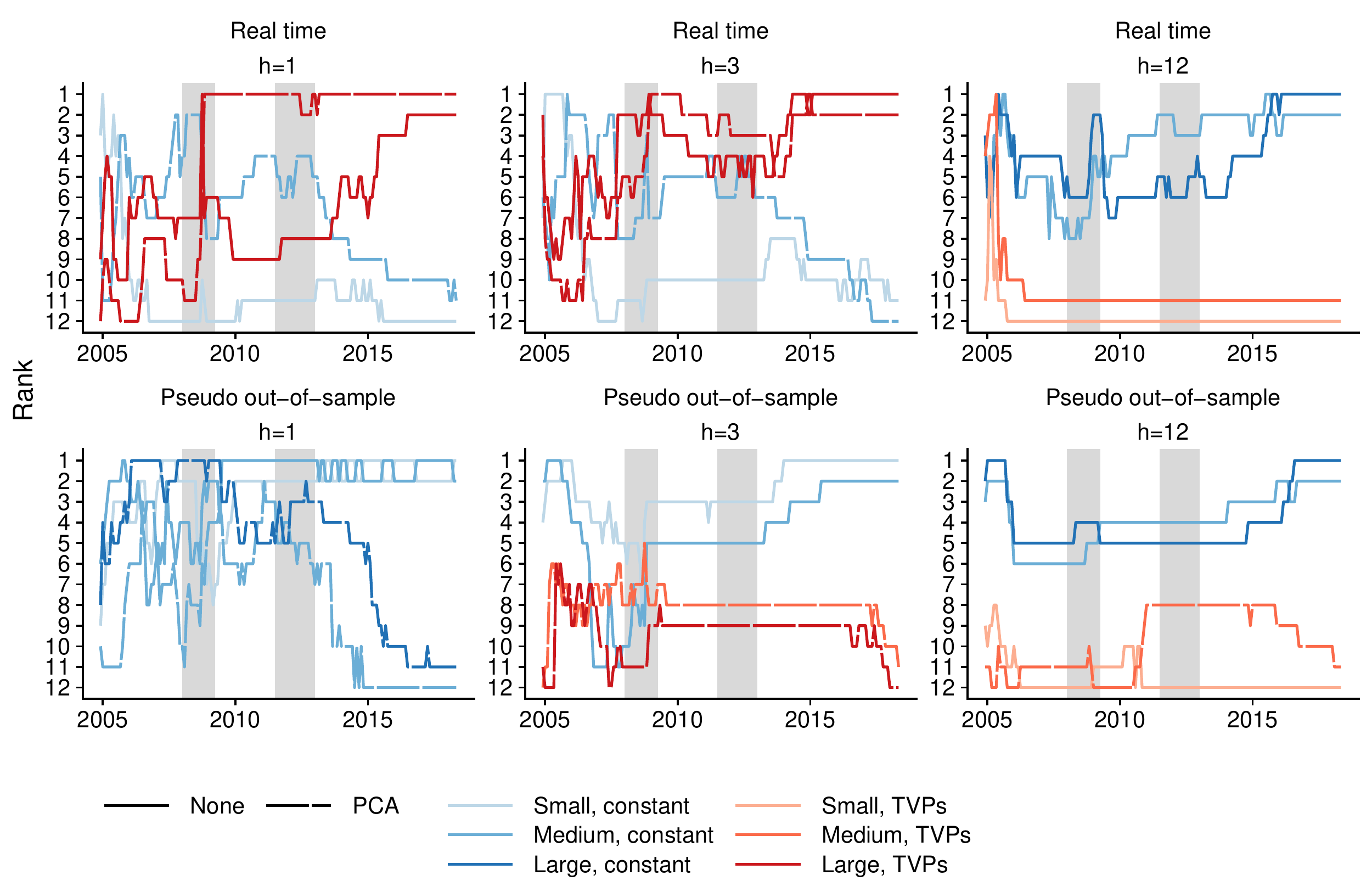}
\end{subfigure}
\end{center}
\caption{Ranking of model specifications over the holdout.}\label{fig:jlps_rank}\vspace*{-0.3cm}
\caption*{\footnotesize\textit{Note}: Ranks are derived from cumulative joint log predictive scores. The figures show the respective two best and worst performing models. The grey shaded areas indicate recessions dated by the NBER Business Cycle Dating Committee (US) and the CEPR Euro Area Business Cycle Dating Committee (EA).}
\end{figure}

We proceed with investigating the relative forecast performance of the competing specifications in real time and based on the pseudo out-of-sample simulations. Figure \ref{fig:jlps_rank} shows the ranking over time, featuring the two best and worst specifications at the end of the holdout for visualization purposes. Starting with the US, panel (a) indicates that small and medium models featuring TVPs without principal components perform well for most of the holdout sample at one-month and one-quarter ahead horizons. Specifications of the baseline model augmented by principal components occupy the lowest ranks. As suggested in the context of our discussion of rank correlations, the best and worst performing models tend to be similar for both information sets in the case of the US. Some differences occur for one-year ahead forecasts. First, we find that the large model featuring TVPs but no principal components outperforms all other specifications for most of the holdout period. Second, models that perform relatively poorly for shorter horizons exhibit significant gains in the Great Recession.

Considering the results for the EA, Fig. \ref{fig:jlps_rank}(b) draws a rather different picture. While in a real time environment the large model augmented with principal components featuring TVPs performs best for most of the holdout at the one-month and one-quarter ahead horizon, smaller models with constant parameters and without principle components dominate all other specifications. This finding translates to values of the rank correlation coefficient close to zero or even negative, as described previously. A striking example is provided by the small constant parameter model without principal components, that performs best for one-month ahead forecasts in pseudo out-of-sample simulations, but is second to last when using real time information. For one-year ahead forecasts, a different picture emerges. The larger constant parameter models without principal components perform best, while the models featuring TVPs are ordered last for most of the holdout for both real time and pseudo out-of-sample contexts.

The final part of this subsection assesses in detail how forecast performance differs across models when adopting a real time and pseudo out-of-sample simulation. For this purpose, each real time model is benchmarked against its complementary specification estimated using the pseudo out-of-sample information set. Relative LPS larger (smaller) than zero indicate that predictive accuracy based on the pseudo out-of-sample information set is superior (inferior). These relative differences can be interpreted as measures capturing the distance between real time and pseudo out-of-sample forecasts, and do not necessarily indicate superior forecast performance, since each real time model is benchmarked against its pseudo out-of-sample counterpart.

\begin{figure}[t]
\begin{subfigure}[b]{\textwidth}
\caption{United States}
\includegraphics[width=\textwidth]{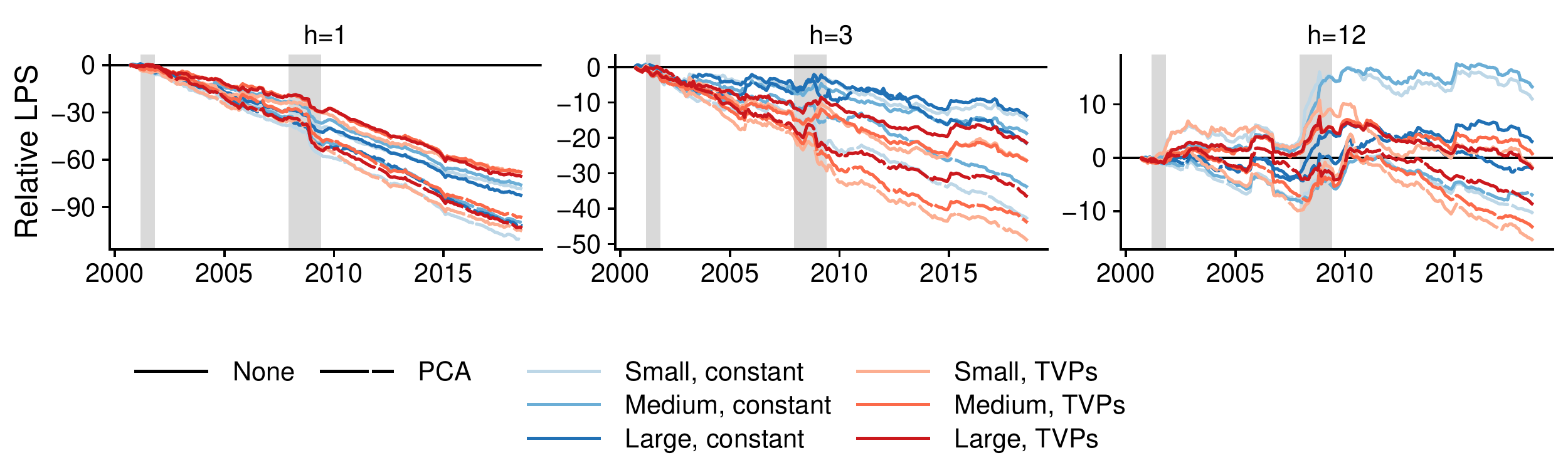}
\end{subfigure}
\begin{subfigure}[b]{\textwidth}
\vspace*{1em}\caption{Euro Area}
\includegraphics[width=\textwidth]{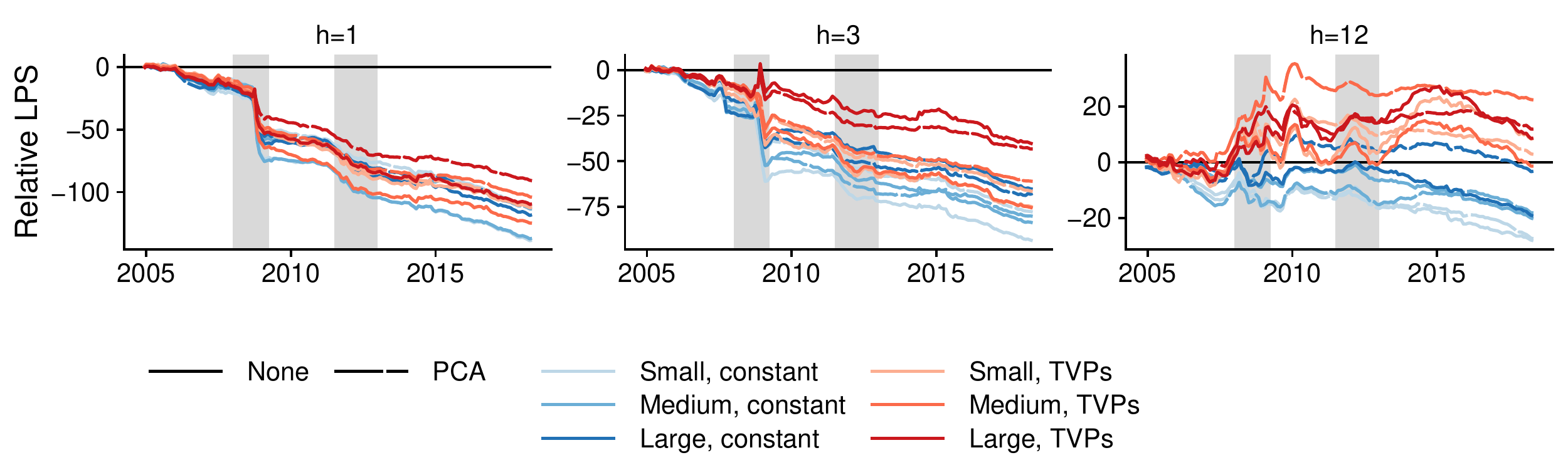}
\end{subfigure}
\caption{Relative cumulative LPSs of real time and pseudo out-of-sample information sets.}\label{fig:jlps_diff}\vspace*{-0.3cm}
\caption*{\footnotesize\textit{Note}: Each real time model is benchmarked against its complementary specification estimated using the pseudo out-of-sample information set. The grey shaded areas indicate recessions dated by the NBER Business Cycle Dating Committee (US) and the CEPR Euro Area Business Cycle Dating Committee (EA).}
\end{figure}

Figure \ref{fig:jlps_diff} shows the resulting relative cumulative LPSs over time. We find that producing forecasts at the one-month and one-quarter ahead horizons using real time information is substantially harder for both the US and EA, indicated by negative relative joint LPSs for all models considered. At the one-month ahead horizon, the distance between the performance metric in real time and pseudo out-of-sample exercise is minimal for medium and large models with TVPs. 

Reconsidering Fig. \ref{fig:jlps_rank}, this implies that such models can be considered comparatively robust to data revisions and imputations, and they tend to perform well in both forecast evaluation contexts. Interestingly, we do not observe clear breaks in relative real time and pseudo out-of-sample performance, but rather consistent differences over time. For one-quarter ahead forecasts, differences are smallest for smaller constant parameter models, and substantial especially for principal component augmented variants with TVPs. At the one-year ahead horizon, we find that some models estimated using real time data outperform their pseudo out-of-sample counterparts, with small to medium constant parameter models showing the most pronounced gains. In contrast to the shorter forecast horizons, clear differences in relative performance emerge during the Great Recession.

A clearer pattern in terms of relative performance using real time and pseudo out-of-sample data is visible for the EA in Fig. \ref{fig:jlps_diff}(b). In particular, for all considered forecast horizons, the constant parameter models show the largest differences between the two forecast evaluation designs. While most specifications estimated using real time data perform worse consistently, we observe deteriorating relative performance measures for one-month and one-quarter ahead forecasts especially during the Great Recession. Similar to the US, the picture is different for one-year ahead forecasts. Again, there are several TVP real time models that outperform their pseudo out-of-sample counterparts.

This discussion of overall forecast performance can be summarized by noting three key observations. First, and perhaps unsurprisingly, producing accurate forecasts in real time is substantially harder than when relying on a pseudo out-of-sample forecast exercise. This is mainly the case at short forecast horizons, and differs for longer horizons. Second, real time and pseudo out-of-sample information sets do not necessarily result in the same relative performance ordering among models. This appears to be a significant problem in the case of the EA. Finally, differences occur both in terms of the specifics of the dataset and release schedule of the vintages, and which forecast horizon is considered.

\subsection{Data vintages and predictions}
Before turning to a more detailed analysis of differences between forecasts in real time and pseudo out-of-sample exercises for different variables and addressing point and density forecasts, we pick the small model with time-varying parameters but without principal components as an example to demonstrate both data features over the different vintages, and discuss differences in the obtained predictive distribution for the real time and pseudo out-of-sample forecast exercise.

\begin{figure}[t]
\begin{subfigure}[b]{\textwidth}
\caption{United States}
\includegraphics[width=\textwidth]{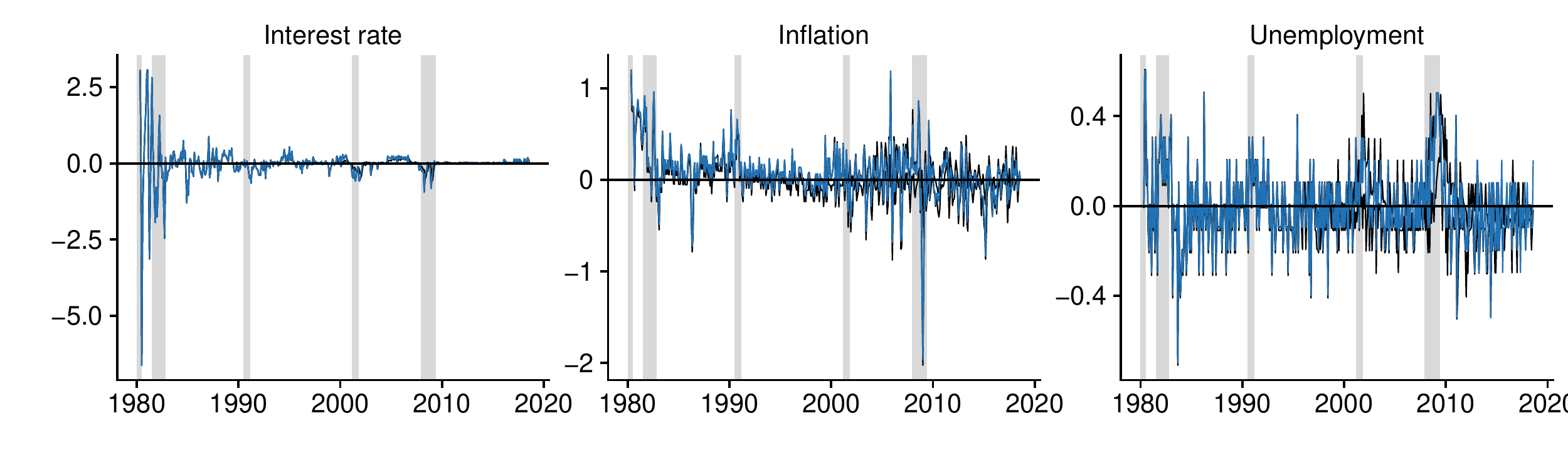}
\end{subfigure}
\begin{subfigure}[b]{\textwidth}
\caption{Euro Area}
\includegraphics[width=\textwidth]{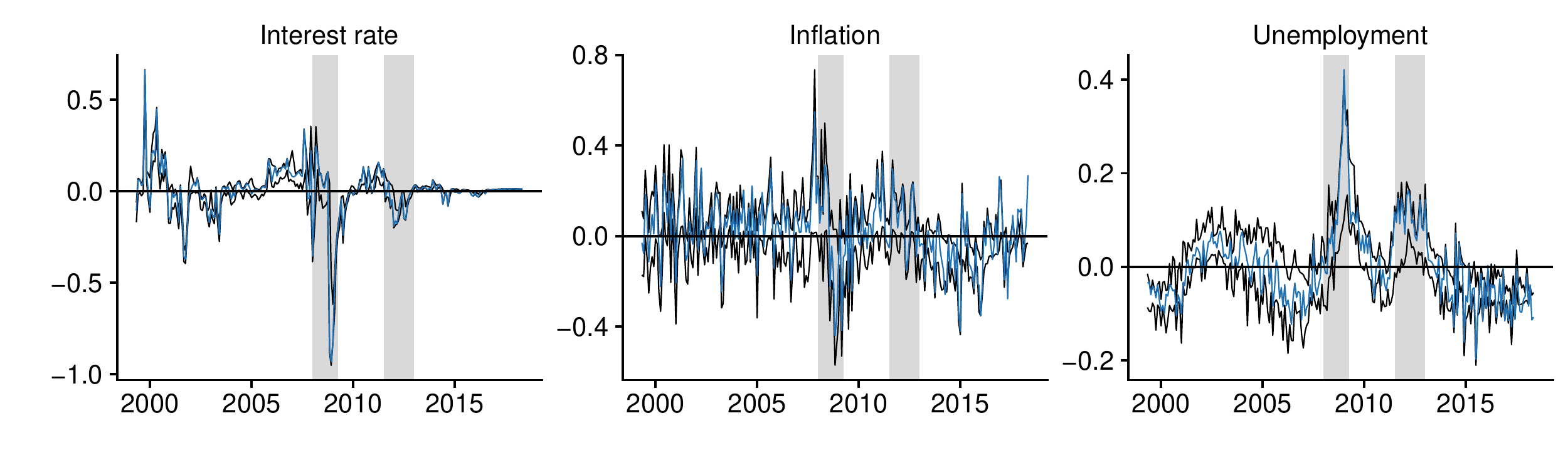}
\end{subfigure}
\caption{Data vintages.}\label{fig:data}\vspace*{-0.3cm}
\caption*{\footnotesize\textit{Note}: The black lines indicate the respective maximum and minimum value at each point in time across all vintages. For imputed data, we present the posterior median estimate of the benchmark specification. The grey shaded areas indicate recessions dated by the NBER Business Cycle Dating Committee (US) and the CEPR Euro Area Business Cycle Dating Committee (EA).}
\end{figure}

Figure \ref{fig:data} shows the full dataset of the focus variables over time, with the blue line marking the final vintage. To visualize the vast amount of data captured in all new releases and the imputed values, we proceed as follows. We use the posterior median of the nowcasted observations and calculate the minimum and maximum value of the respective series per period (black lines) across all vintages. The figure thus indicates both data revisions and uncertainty surrounding imputed values.

We start with the US in panel (a). Since historical vintage data starts only in late 1998, data imputation only plays a role past this date. Interestingly, differences across vintages are also visible prior to this date, implying that earlier data are also subject to revisions. This notion is especially important for inflation. After 1998, the main differences originate from the nowcasting scheme, with largest deviations observable during recessionary episodes for inflation and unemployment. Interest rates, on the other hand, are comparatively unaffected by revisions and data imputation.

The same is true for the EA. Data revisions and imputations in the EA, shown in Fig. \ref{fig:data}(b), play only a minor role for interest rates, evidenced by the narrow bounds surrounding the final vintage. For inflation and unemployment, on the other hand, we find substantial differences of the vintage data over time. The largest changes are visible during the Great Recession starting in 2008, and the European debt crisis between 2011 and 2013. This observation is particularly striking for unemployment, with negative deviations up to 0.2 percentage points between nowcasts and the final series.

\begin{figure}[t]
\begin{subfigure}[b]{\textwidth}
\caption{United States}
\includegraphics[width=\textwidth]{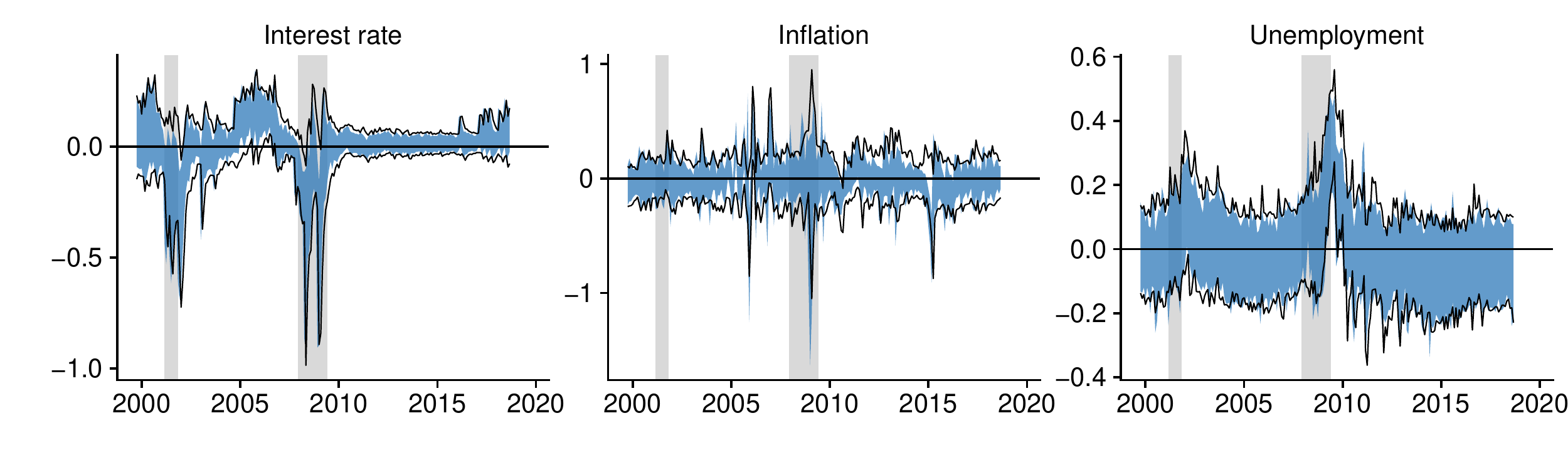}
\end{subfigure}
\begin{subfigure}[b]{\textwidth}
\caption{Euro Area}
\includegraphics[width=\textwidth]{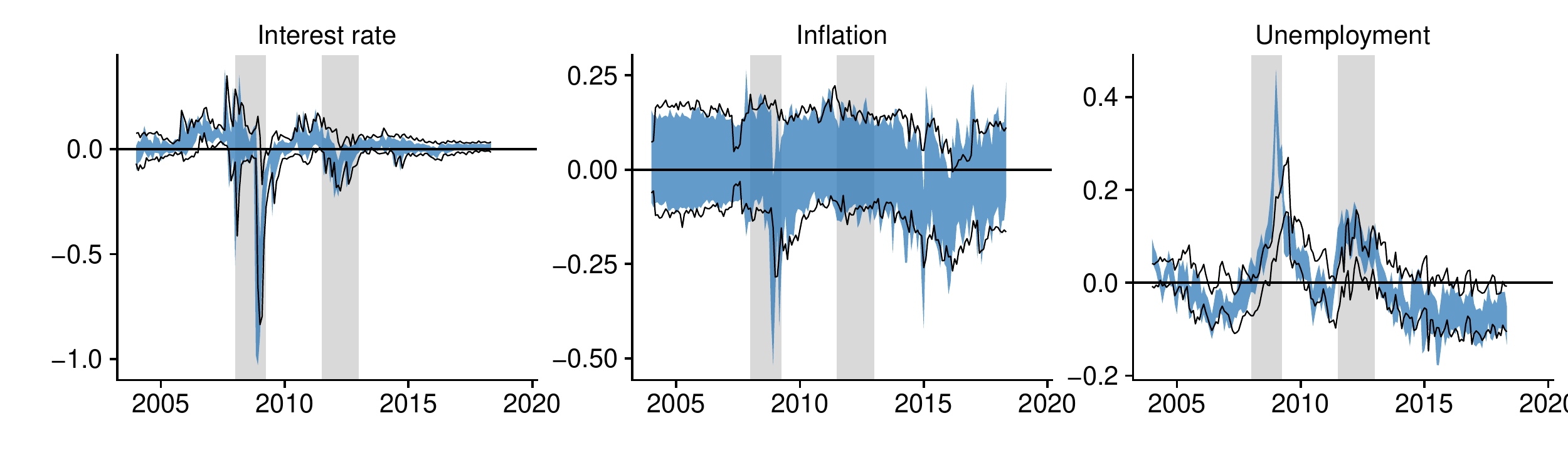}
\end{subfigure}
\caption{One-step ahead predictive distributions for the focus variables.}\label{fig:preds}\vspace*{-0.3cm}
\caption*{\footnotesize\textit{Note}: The black lines capture the $68$ percent posterior coverage interval for the real time data, the blue shaded area refers to the corresponding pseudo out-of-sample exercise. The blue line is the realized series. The grey shaded areas indicate recessions dated by the NBER Business Cycle Dating Committee (US) and the CEPR Euro Area Business Cycle Dating Committee (EA).}
\end{figure}

To assess out-of-sample features both in a real time and pseudo out-of-sample exercise, we depict the respective one-step ahead predictive distributions alongside the $68$ percent posterior coverage interval in Fig. \ref{fig:preds}. The blue shaded area captures pseudo out-of-sample forecasts and the black lines are based on real time information sets.

In Fig. \ref{fig:preds}(a), it is clearly visible that taking a real time perspective yields wider credible sets for the predictive distribution for the US, originating from the need of nowcasting some of the involved quantities. Interestingly, even though interest rates are less affected by data revisions, we also detect differences in the predictive distribution stemming from missing values in the other variables. In general, the benchmark model performs well for forecasting, with the predictive intervals covering the realized series in most cases. Exceptions are various months during recessions, where some values lie outside the credible set.

For the EA (see Fig. \ref{fig:preds}(b)), a key notion again is that the posterior distribution for predictions based on real time data are wider than for pseudo out-of-sample forecasts for all focus variables, even more so than for the US. This increased uncertainty surrounding forecasts stems from both the imputed values and data revisions. While forecasts track the actual evolution of interest rates satisfactorily, we observe that some peaks and troughs of inflation lie outside the bounds of the credible set. An interesting finding for unemployment is that the model consistently predicts lower unemployment rates in real time during the Great Recession, while for the pseudo out-of-sample exercise, the posterior more accurately tracks the evolution of the realized series. The same is visible during the European debt crisis, albeit to a lesser extent.

\subsection{Forecasts for the United States}\label{sec:forecasts_us}
In Sections \ref{sec:forecasts_us} and \ref{sec:forecasts_ea}, we provide a more detailed discussion of differences in forecasts when adopting a real time perspective individually for the US and the EA. In particular, we assess specific findings for point and density forecasts across focus variables and horizons. Additional results to identify the best performing models and differences between real time and pseudo out-of-sample contexts are provided in Appendix \ref{app:C}.

Figure \ref{fig:US_corr} shows Kendall's rank correlation coefficient $\tau$ for the forecast performance at different horizons and across variable types based on both point and density forecasts. For point forecasts, we assign ranks based on minimal cumulative absolute forecast errors (FEs), while ranks for predictive densities are constructed from cumulative marginal LPSs in descending order. A striking observation is that rank correlations for density forecasts and point forecasts exhibit different patterns over time across most variable types and forecast horizons. 

\begin{figure}[!t]
\includegraphics[width=\textwidth]{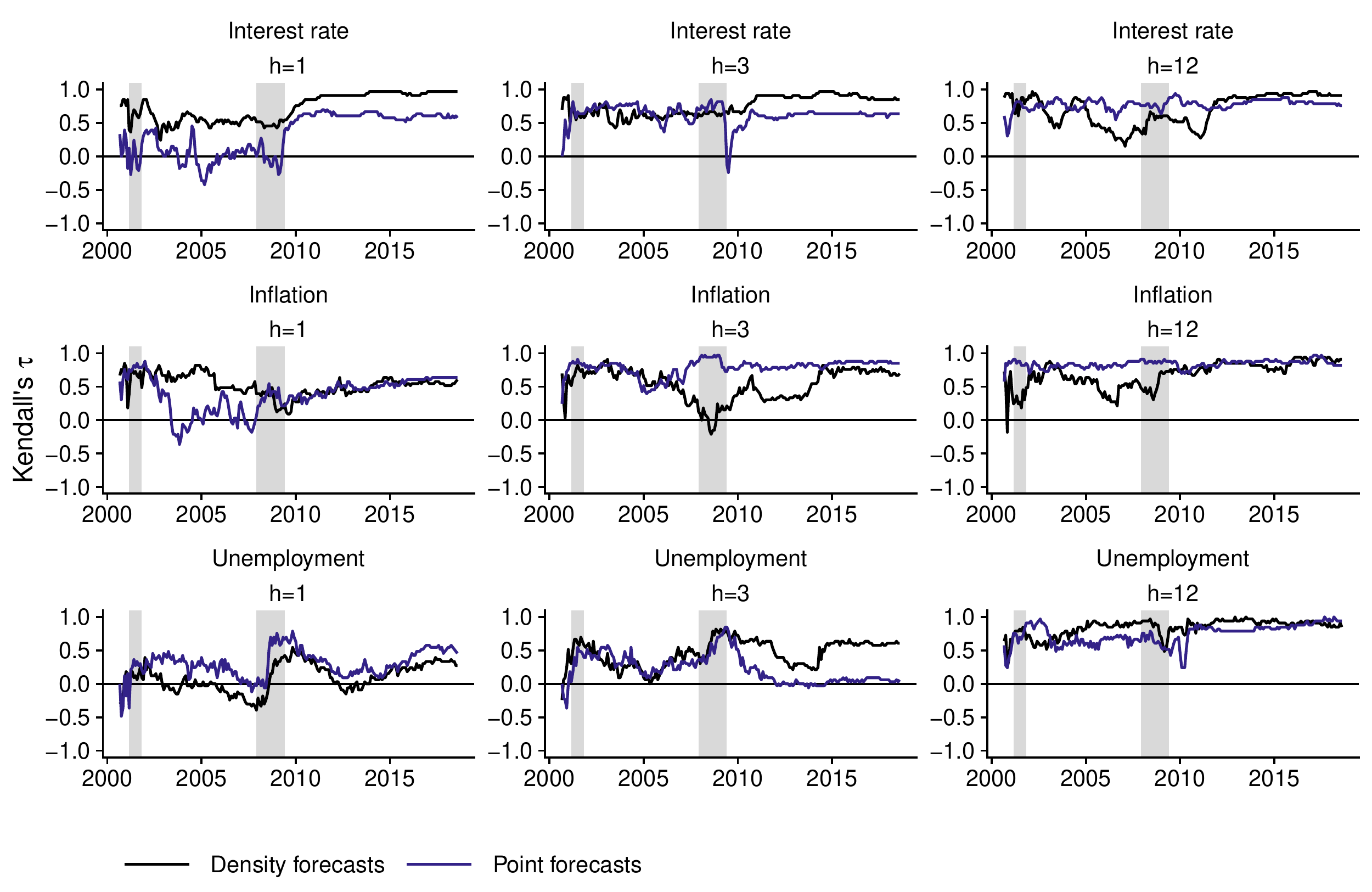}
\caption{Rank correlation of different specifications between real time and pseudo out-of-sample forecasts for the US.}\label{fig:US_corr}\vspace*{-0.3cm}
\caption*{\footnotesize\textit{Note}: Ranks are derived from cumulative log predictive scores (Density forecasts) and cumulative absolute forecast errors (Point forecasts). The grey shaded areas indicate recessions dated by the NBER Business Cycle Dating Committee.}
\end{figure}

At the one-month ahead horizon, we find that the relative performance in terms of density forecasts is similar in real time and pseudo out-of-sample contexts for interest rates and inflation, with the rank correlation coefficient close to one for most of the holdout. Here, especially models with TVPs appear to perform well. For unemployment, there is essentially no correlation in rank orders until the Great Recession. During the recovery period, we observe a slightly higher agreement in terms of relative forecast performance for density forecasts. However, this robustness to data revisions and imputations fades slowly in the later part of the holdout, fluctuating around $0.3$ at the end of the considered period. In a real time context, introducing additional information via principal components tends to pay off.

Conversely, for point forecasts, we find almost no association in the rank ordering until the Great Recession for all variables. Afterwards, coinciding with the period when the zero lower bound was reached, the relative orderings tend to agree for interest rates, and to a slightly lesser extent for inflation. The rank correlation coefficients over time visibly show comovement for point and density forecasts of unemployment, albeit the correlation is higher for point forecasts. It is noteworthy that models that perform well for density forecasts do not necessarily perform well for point forecasts, especially for inflation. Interestingly, the large constant parameter model featuring principal components produces accurate point and density forecasts in real time and pseudo out-of-sample contexts for unemployment.

For one-quarter ahead forecasts of interest rates, we find that rank orderings consistently agree for point and density forecasts, although we observe a brief period of lower correlations just after the Great Recession in terms of point forecasts. The relative performance for point forecasts of inflation is rather stable and strongly positively correlated for real time and pseudo out-of-sample exercises. Conversely, we find that although the performance ordering agrees early in the sample for one-quarter ahead density forecasts of inflation, the rank correlation coefficient is zero during the Great Recession. After that, we again observe a clearly positive correlation between real time and pseudo out-of-sample relative performance ranking. Here, TVPs seem to be crucial to provide accurate density forecasts, while best point forecasts are achieved by small scale constant parameter models. 

Turning to one-quarter ahead forecasts of unemployment, we observe fluctuating correlations between almost zero and $0.8$, with point and density rank correlations exhibiting a substantial degree of comovement. This relationship decouples during the Great Recession, with high correlations in terms of densities, and zero association between real time and pseudo out-of-sample rank ordering for point forecasts afterwards. Larger TVP specifications without principal components appear to produce the best point and density forecasts for both real time and pseudo out-of-sample simulations, on average.

One-year ahead forecasts appear to be remarkably robust in terms of relative performance orderings. For point forecasts, the rank correlation coefficient is close to one for all three focus variables for most of the holdout, indicating that real time and pseudo out-of-sample exercises produce almost identical rankings of model performance. For density forecasts, we observe a similar situation, although we identify lengthy periods before $2010$ with correlations around $0.5$, especially for the interest rate and inflation. In general, we find that constant parameter models produce superior forecasts at longer horizons, except for density forecasts of inflation.

\subsection{Forecasts for the Euro Area}\label{sec:forecasts_ea}
\begin{figure}[t]
\includegraphics[width=\textwidth]{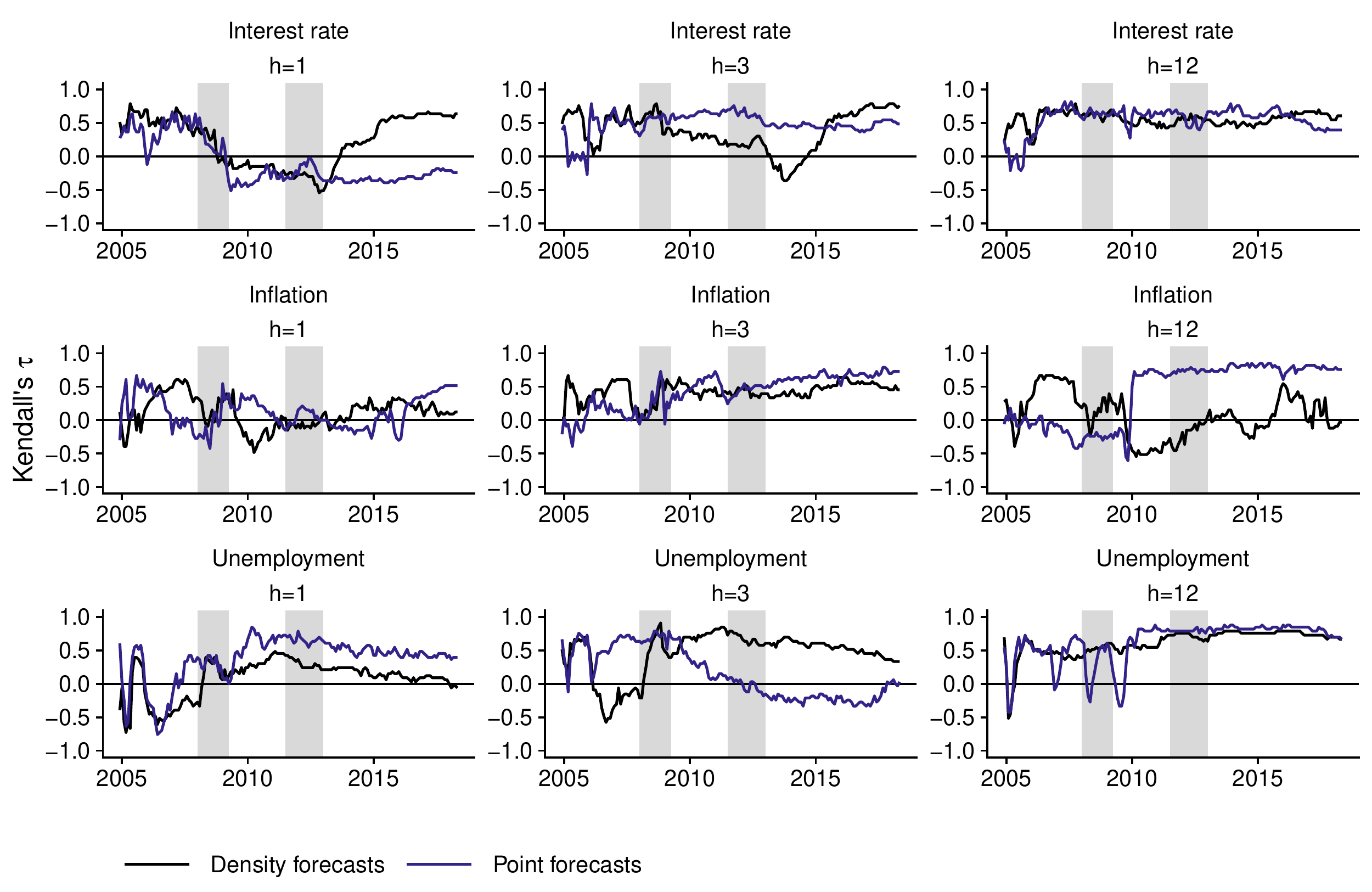}
\caption{Rank correlation of different specifications between real time and pseudo out-of-sample forecasts for the EA.}\label{fig:EA_corr}\vspace*{-0.3cm}
\caption*{\footnotesize\textit{Note}: Ranks are derived from cumulative log predictive scores (Density forecasts) and cumulative absolute forecast errors (Point forecasts). The grey shaded areas indicate recessions dated by the CEPR Euro Area Business Cycle Dating Committee.}
\end{figure}

Variable specific differences for relative orderings of point and density forecast performance in the EA are shown in Fig. \ref{fig:EA_corr}. As suggested in Section \ref{sec:results_overall}, taking a real time perspective is even more consequential for relative forecast performance for the EA than the US. This observation is true for both point and density forecasts.

Starting with the one-month ahead horizon, point and density forecasts for interest rates exhibit a similar ordering for real time and pseudo out-of-sample simulations until the Great Recession. Overall, TVPs appear to improve forecast accuracy. In the period between the two recessions, we observe negative rank correlations, indicating that using real time information reverses the ordering of the considered models in terms of their relative forecast accuracy. In the period after the European debt crisis when the zero lower bound was reached, density forecasts are again similar for real time and pseudo out-of-sample information, while this is not the case for point forecasts. This is mainly due to the relatively poor performance of smaller constant parameter models augmented with principal components in the pseudo out-of-sample simulation, specifications that appear to perform satisfactorily in real time.

A different picture emerges for short-horizon inflation forecasts, with no clear patterns in terms of rank correlation, albeit changes in relative performance orderings appear to occur especially during and between the two recessions. While larger models with TVPs and without principal components perform well for density forecasts, small to medium size models with constant parameters indicate superior performance for point forecasts. 

For unemployment, similar as in the US, we detect comovements in correlations over time for point and density forecasts. Point forecast accuracy orderings between real time and pseudo out-of-sample contexts tend to be concordant after the Great Recession. Density forecasts, however, exhibit substantially lower and decreasing correlation measures over time, with values close to zero at the end of the holdout. Large fluctuations for inflation and unemployment early in the sample may be explained by the relatively short initial estimation period. No clear best performing specification in real time and pseudo out-of-sample contexts can be identified, although constant parameter models appear to be the most robust.

Interest rate density forecasts for the one-quarter horizon show similar concordance as in the one-month ahead case, again with gains for models featuring TVPs. Point forecasts mostly agree on model ordering, different to the shorter horizon. Principal components seem to improve forecast accuracy in both the TVP and constant parameter cases. For inflation, we observe higher correlations especially after the Great Recession, indicating that pseudo out-of-sample exercises establish a similar model ordering when compared to relying on real time information. Here, TVP models are superior for density forecasts, while constant parameter models are superior for point predictions.

Interestingly, for unemployment, correlation measures show a similar path as for the US. Early in the holdout, the measure fluctuates substantially, increasing during the Great Recession. This implies that for density forecasts, both real time and pseudo out-of-sample simulations produce similar orderings of relative performance. The small constant parameter model featuring principal components performs relatively similar in both evaluation contexts, while TVP models are ranked last in both cases. Point forecast correlations deteriorate afterwards, and are close to zero or even negative after the first recession.

Long-horizon forecasts exhibit different patterns for rank correlations. For interest rates, consequences for the relative performance orderings derived from real time and pseudo out-of-sample simulations are present but muted, indicated by a coefficient that fluctuates just above $0.5$ for most of the holdout. TVPs play an important role for point and density forecasts both using real time and truncated vintage data. Inflation forecast performance orderings are discordant for point forecasts early in the sample, however, tend to agree on the rank order after the Great Recession. Small scale constant parameter models featuring principal components perform well. Density forecasts, on the other hand, tend to be ordered similarly early in the sample. This changes after the Great Recession, with negative $\tau$ suggesting inverse performance orderings of the competing models. At the end of the holdout sample, the association between performance orderings are close to zero. The best performing model in real time here is the large TVP-VAR without principal components, while the medium scale constant parameter model works best in the pseudo out-of-sample exercise.

For unemployment, we find consistently high values of correlations throughout the holdout period in terms of density forecasts. This implies that pseudo out-of-sample evaluation is sufficient to establish a useful ordering of model performance. Regarding point forecasts, we observe some periods until $2010$ where substantial rank changes occur. From $2010$ onwards, Kendall's rank correlation is close to one, again indicating that relying on real time information does not affect relative performance orderings. Here, constant parameter specifications appear to be superior for point and density forecasts in both information sets.

\section{Closing remarks}\label{sec:conclusions}
In this paper, we systematically assess differences in forecast performance between a set of differently sized models using pseudo out-of-sample versus real time simulations. We rely on constant and TVP-VAR models with SV, equipped with a global-local shrinkage prior. We also consider variants augmented by principal components to capture high-dimensional information from available datasets, and discuss imputing missing values. Our results suggest differences in the relative ordering of model performance for point and density forecasts. No clearly superior specification for the US or the EA across variable types and forecast horizons can be identified, although larger models featuring TVPs appear to be affected the least by missing values and data revisions. This finding suggests that pseudo out-of-sample simulations are insufficient to establish a clear ordering of relative model performance, and attention in the development of new methods should always be paid to real time features of data in order to be of value for forecasters in central banks and governments.

\small{\setstretch{0.85}
\addcontentsline{toc}{section}{References}
\bibliographystyle{custom.bst}
\bibliography{lit}}

\begin{appendices}\crefalias{section}{appsec}
\setcounter{equation}{0}
\renewcommand\theequation{A.\arabic{equation}}
\section{Econometric appendix}\label{app:A}
\subsection{The horseshoe prior}\label{app:horseshoe}
For simplicity, we construct a $2K_{i}\times1$-dimensional vector containing the constant part of the coefficients, covariances, and the associated state innovation variances for equation $i$ of the VAR, $\bm{b}_i = \left(\bm{\beta}_{i0}',\{v_{il}\}_{l=1}^{K_i}\right)'$, and denote the $j$th element of this vector by $b_{ij}~(j=1,\hdots,2K_i)$.\footnote{For the constant parameter specifications, $\bm{b}_i = \bm{\beta}_{i0}$ of dimension $K_i\times1$, where the posteriors for the horseshoe prior have to account for this change in the number of free parameters.} As in \citet{HKO}, we establish the horseshoe prior of \citet{carvalho2010horseshoe} based on the auxiliary variable representation in \citet{makalic2015simple},
\begin{equation*}
b_{ij}|\lambda_i,\psi_{ij}\sim\mathcal{N}(0,\psi_{ij}\lambda_i), \quad \psi_{ij}\sim\mathcal{G}^{-1}\left(1/2,\zeta_{ij}^{-1}\right), \quad \lambda_{i}\sim\mathcal{G}^{-1}(1/2,\varphi_i^{-1}),
\end{equation*}
with $\mathcal{G}^{-1}$ referring to the inverse Gamma distribution. The auxiliary variables are assigned the priors $\varphi_i\sim\mathcal{G}^{-1}(1/2,1)$ and $\zeta_{ij}\sim\mathcal{G}^{-1}(1/2,1)$.

\subsection{Priors for stochastic volatilities}
For the stochastic volatilities, we rely on the R-package \texttt{stochvol} and employ the priors proposed in \citet{KASTNER2014408}. Specifically, we use a Gaussian prior for the level of the log-volatilities $\mu_{i}\sim\mathcal{N}(0,100)$. For the autoregressive coefficient $\phi_i\in(-1,1)$, a transformed Beta distribution is used, $(\phi_i+1)/2\sim\mathcal{B}(25,5)$, while a Gaussian prior on the signed square root of the innovation variances is employed, $\pm\sqrt{\varsigma_i}\sim\mathcal{N}(0,1)$. The prior on the initial state is based on the stationary distribution of the process, $\sigma_{i0}|\mu_i,\varsigma_i,\phi_i\sim\mathcal{N}\left(\mu_i,\varsigma_i^2/(1-\phi_i^2)\right)$.

\subsection{Posterior simulation and algorithm}
\begin{enumerate}[leftmargin=1.5em,itemsep=0em]
\item Given a draw of the full history of the time-varying part of the model coefficients and the stochastic volatilities, the VAR coefficients, covariances and associated state innovation variances in $\bm{b}_i$ can be drawn jointly (on an equation-by-equation basis) from their multivariate Gaussian conditional posterior distribution with standard moments for linear regression models \citep[see, for instance,][]{koop2003bayesian}. For the exact formulae of the moments, refer to \citet{feldkircher2017sophisticated}.
\item Conditional on all other model parameters, a forward-filtering backward-sampling \citep[FFBS,][]{doi:10.1093/biomet/81.3.541,doi:10.1111/j.1467-9892.1994.tb00184.x} algorithm is employed again on an equation-by-equation basis for drawing the full history of the TVPs. This step is omitted for the constant parameter models.
\item The posteriors associated with the horseshoe prior established in Appendix \ref{app:horseshoe} are
\begin{align*}
\psi_{ij}|b_{ij},\lambda_i,\zeta_{ij} &\sim\mathcal{G}^{-1}\left(1,\zeta_{ij}^{-1}+\frac{b_{ij}^2}{2\lambda_i}\right)\\
\lambda_{i}|b_{ij},\psi_i,\varphi_i &\sim\mathcal{G}^{-1}\left(\frac{2K_i+1}{2},\varphi_i^{-1}+\frac{1}{2}\sum_{j=1}^{2K_i}b_{ij}^2\psi_{ij}^{-1}\right)
\end{align*}
for the global and local shrinkage parameters. The conditional posterior distributions for the auxiliary variables are $\zeta_{ij}|\psi_{ij} \sim\mathcal{G}^{-1}\left(1,1+\psi_{ij}^{-2}\right)$ and $\varphi_{i}|\lambda_i \sim\mathcal{G}^{-1}\left(1,1+\lambda_i^{-2}\right)$.
\item Sampling of the stochastic volatilities and associated parameters is carried out using the algorithm in \citet{KASTNER2014408} and its implementation in the R-package \texttt{stochvol}.
\item Depending on the forecast exercise, we impute missing values in the endogenous variables using the quantities discussed in Section \ref{sec:nowcasting}.
\end{enumerate}
We use a total number of $6,000$ draws and discard the initial $2,000$ draws as burn-in. We consider each second draw to obtain a set of $S=2,000$ posterior draws. These draws can be used for calculating the predictive distribution ex post, and allow for simulating point and density forecasts that are used to obtain the performance metrics.

\subsection{Producing and evaluating forecasts}
For calculating iterated $h$-step ahead forecasts, we use the VAR coefficients $\bm{A}_T$ and covariance matrix $\bm{\Omega}_T$ at time $T$, let $\bm{Y}_T=(\bm{y}_T',\bm{y}_{T-1}',\hdots,\bm{y}_{T-P+1}')'$, $\bm{\nu}_T = (\bm{\epsilon}_T,\bm{0}_M',\hdots,\bm{0}_M')'$ and denote the respective companion and covariance matrix by
\begin{equation*}
\bm{\tilde{A}} = \begin{bmatrix}
\bm{A}_{1T} & \bm{A}_{2T} & \hdots & \bm{A}_{P-1,T} & \bm{A}_{PT}\\
\bm{I}_M    & \bm{0}_M    & \hdots & \bm{0}_M       & \bm{0}_M\\
\bm{0}_M    & \bm{I}_M    & \hdots & \bm{0}_M       & \bm{0}_M\\
\vdots      &             & \ddots &                & \vdots\\
\bm{0}_M    & \hdots      & \hdots & \bm{I}_M       & \bm{0}_M
\end{bmatrix}, 
\quad \text{Var}(\bm{\nu}_T) = \bm{\tilde{\Omega}} = \begin{bmatrix}
\bm{\Omega}_T & \bm{0}_M  & \hdots & \bm{0}_M\\
\bm{0}_M      & \bm{0}_M  &        & \vdots\\
\vdots        &           & \ddots & \vdots\\ 
\bm{0}_M      & \hdots    & \hdots & \bm{0}_M
\end{bmatrix}.
\end{equation*}
The moments for the distribution of the $h$-step ahead forecasts can be derived by observing
\begin{align*}
\bm{\mu}_{T+h}    &= \text{E}(\bm{Y}_{T+h}) = \bm{\tilde{A}}^h \bm{Y}_{T}\\
\bm{\Sigma}_{T+h} &= \text{E}(\bm{Y}_{T+h}\bm{Y}_{T+h}') = \sum_{i=0}^{h-1} \bm{\tilde{A}}^i \bm{\tilde{\Omega}} \bm{\tilde{A}}^{i}{'}.
\end{align*}
The forecast $\bm{y}_{T+h}^{(f)}$ arises from a Gaussian distribution,
\begin{equation}
\bm{\tilde{y}}_{T+h}^{(f)} \sim \mathcal{N}\left(\bm{\tilde{\mu}}_{T+h}, \bm{\tilde{\Sigma}}_{T+h}\right),\label{eq:pred}
\end{equation}
with moments $\bm{\tilde\mu}_{T+h} = \bm{\mu}_{T+h}^{[1:M]}$ and $\bm{\tilde{\Sigma}}_{T+h} = \bm{{\Sigma}}_{T+h}^{[1:M]}$, whereby the superscript $[1:M]$ indicates selecting the first $M$ elements or $M\times M$ block in the respective vector or matrix. Taking account of standardizing the series, the non-standardized forecast is $\bm{y}_{T+h}^{(f)} = \left(\bm{\tilde{y}}_{T+h}^{(f)}\odot \bm{s}_y\right)+\bm{m}_y$ where $\odot$ denotes element-wise multiplication. We assess the performance of both point and density forecasts. For measuring the precision of first moment forecasts, we rely on the root mean squared error (RMSE), while second moment forecasts are evaluated by the log predictive likelihood (LPL) constructed from the full predictive distribution. 

To obtain a point forecast, we store draws from \autoref{eq:pred} for all iterations of the algorithm and refer to the mean of this distribution for series $i$ by $\hat{y}_{i,t+h}^{(f)}$. Let $T_H$ indicate the length of the holdout sample, $\bm{y}_{T+h}^{(r)}$ are the corresponding realized values, then RMSEs for the $i$th variable (for $i=1,\hdots,M$) are defined as
\begin{equation*}
\text{RMSE}_{ih} = \sqrt{\frac{1}{T_H-T} \sum_{t=T}^{T_H} \left(y_{i,t+h}^{(r)}-\hat{y}_{i,t+h}^{(f)}\right)^2}
\end{equation*}
This measure captures the average deviation of the forecast from realizations over the holdout period. Absolute forecast errors (FEs) by period are closely related, and are calculated as 
\begin{equation*}
\text{FE}_{i,t+h} = |y_{i,t+h}^{(r)}-\hat{y}_{i,t+h}^{(f)}|. 
\end{equation*}

A key aspect in forecasting, however, is how precise the forecasts are, which is also reflected in the second moment of the predicive distribution. The marginal predictive likelihood (MPL) for each period of the holdout at time $t$ for the $h$-step ahead forecast of variable $i$ is given by 
\begin{equation}
\text{MPL}_{i,t+h} = f_{\mathcal{N}}\left(y_{i,t+h}^{(r)}|m_{yi}+s_{yi}\tilde{\mu}_{i,t+h},s_{yi}^2\bm{\tilde\Sigma}_{ii,t+h}\right),\label{eq:lpl}
\end{equation}
where $f_\mathcal{N}$ denotes the probability density function of the normal distribution with the realized value being evaluated in the predictive distribution. Note that if the data are not standardized a priori, $m_{yi}=0$, $s_{yi}=1$, and the moments in \autoref{eq:lpl} correspond to \autoref{eq:pred}. This distribution can be evaluated for each iteration of the algorithm, taking into account both intrinsic uncertainty from the error term and estimation uncertainty for parameters. Following \citet{geweke2010comparing}, the LPLs are defined as
\begin{equation*}
\text{LPL}_{i,t+h} = \log\left(\frac{1}{S}\sum_{s=1}^S \text{MPL}_{i,t+h}^{(s)}\right)
\end{equation*}
Here, $s=1,\hdots,S$ refers to the iterations of the sampler with a total number of $S$ draws from the posterior distribution. To assess overall forecast performance of a model, one may also evaluate the joint predictive likelihood (JPL):
\begin{equation}
\text{JPL}_{t+h} = f_{\mathcal{N}}\left(\bm{y}_{t+h}^{(r)}|\bm{m}_{y}+\bm{s}_{y}\odot\bm{\tilde{\mu}}_{t+h},\bm{\tilde S}_{t+h}\right).\label{eq:tpl}
\end{equation}
The non-normalized predictive covariance matrix is constructed using a Cholesky factorization of $\bm{\tilde\Sigma}_{t+h}=\bm{L}_{t+h}\bm{L}_{t+h}'$ and the normalizing standard deviations in $\bm{S}_y = \text{diag}(s_{y1},\hdots,s_{yM})$ to yield $\bm{\tilde S}_{t+h} = (\bm{S}_y\bm{L}_{t+h})(\bm{S}_y\bm{L}_{t+h})'$. Note that we evaluate the joint predictive likelihood solely for the focus variables, which amounts to subsetting the vectors and matrices in \autoref{eq:tpl} to the corresponding elements. The joint log predictive likelihood is
\begin{equation*}
\text{joint LPL}_{t+h} = \log\left(\frac{1}{S}\sum_{s=1}^S \text{JPL}_{t+h}^{(s)}\right).
\end{equation*}

\setcounter{equation}{0}
\renewcommand\theequation{B.\arabic{equation}}
\section{Data appendix}\label{app:B}
We obtain data for the US and the EA to reflect similar information sets. All series (when applicable) are seasonally adjusted, see also Section \ref{sec:data}. To avoid confounding information sets, all adjustments such as for seasonality, transformations to stationarity and standardizing the data are applied to the vintages independently. The transformation of a series $x_t$ for obtaining stationarity are: (1) no transformation, (2) $\Delta x_t$, (5) $\Delta \log(x_t)$, (6) $\Delta^2 \log(x_t)$ with $\Delta^i$ indicating $i$th differences. The transformations for the US dataset are based on suggestions in \citet{doi:10.1080/07350015.2015.1086655}, while for the EA we use corresponding transformations for the respective series.

\subsection{FRED-MD data}
The Federal Reserve Economic Data (FRED) is maintained by the Federal Reserve Bank of St. Louis and available for download at \href{https://research.stlouisfed.org/}{research.stlouisfed.org}. All series are on a monthly frequency, with some of them starting in 1959:01. Data vintages are available from 1999:08. The final vintage contains $129$ series -- we preselect the $99$ series used for this paper, as mentioned in Section \ref{sec:data}, based on consistent availability of all historical data. Moreover, some variables are dropped due to substantial publication lags exceeding six months \citep[see also][]{doi:10.1080/07350015.2015.1086655}. Base year changes are accounted for by normalizing the respective series to a unique basis.

\begin{tiny}
\vspace*{0.5cm}
\begin{longtable}{p{2cm}p{6.5cm}p{0.5cm}C{1.2cm}C{1.2cm}C{1.2cm}}
\caption{Variables in the FRED-MD dataset.}\\[-1em]
\toprule
\textbf{Abbreviation} & \textbf{Description} & \textbf{Tc} & \textbf{Small} & \textbf{Medium} & \textbf{Large} \\ 
\midrule\endfirsthead
\textbf{Abbreviation} & \textbf{Description} & \textbf{Tc} & \textbf{Small} & \textbf{Medium} & \textbf{Large} \\ 
\midrule
\endhead

\midrule
\endfoot
\multicolumn{6}{p{15cm}}{\textit{Notes}: The dataset described in \citet{doi:10.1080/07350015.2015.1086655} is available for download at \href{https://research.stlouisfed.org/}{research.stlouisfed.org} (FRED-MD). Column \textbf{Tc} indicates the transformation of a series $x_t$ for obtaining stationarity: (1) no transformation, (2) $\Delta x_t$, (5) $\Delta \log(x_t)$, (6) $\Delta^2 \log(x_t)$ with $\Delta^i$ indicating $i$th differences. Columns Small, Medium and Large refer to different model sizes discussed in Section \ref{sec:data}.}
\endlastfoot
  UNRATE & Civilian Unemployment Rate & 2 & x & x & x \\ 
  FEDFUNDS & Effective Federal Funds Rate & 2 & x & x & x \\ 
  CPIAUCSL & CPI: All Items & 5 & x & x & x \\ 
  INDPRO & IP Index  & 5 &  & x & x \\ 
  S\&P 500 & S\&P's Common Stock Price Index: Composite & 5 &  & x & x \\ 
  GS10 & 10-Year Treasury Rate & 2 &  & x & x \\ 
  M2SL & M2 Money Stock & 6 &  &  & x \\ 
  BUSLOANS & Commercial and Industrial Loans  & 6 &  &  & x \\ 
  T10YFFM & 10-Year Treasury C Minus FEDFUNDS & 1 &  &  & x \\ 
  EXUSUKx & U.S.--UK Foreign Exchange Rate & 5 &  &  & x \\ 
  OILPRICEx & Crude Oil, spliced WTI and Cushing & 6 &  &  & x \\ 
  \midrule
  RPI & Real personal income   & 5 &  &  &  \\ 
  W875RX1 & Real personal income ex transfer receipts & 5 &  &  &  \\ 
  CMRMTSPLx & Real Manu. and TradeIndustries Sales & 5 &  &  &  \\ 
  RETAILx & Retail and Food Services Sales  & 5 &  &  &  \\ 
  IPFPNSS & IP: Final Products & 5 &  &  &  \\ 
  IPFINAL & IP: Final Products (Market Group) & 5 &  &  &  \\ 
  IPCONGD & IP: Consumer Goods & 5 &  &  &  \\ 
  IPMAT & IP: Materials  & 5 &  &  &  \\ 
  IPMANSICS & IP: Manufacturing (SIC) & 5 &  &  &  \\ 
  CUMFNS & Capacity Utilization: Manufacturing & 2 &  &  &  \\ 
  CLF16OV & Civilian Labor Force & 5 &  &  &  \\ 
  CE16OV & Civilian Employment  & 5 &  &  &  \\ 
  UEMPMEAN & Average Duration of Unemployment (Weeks) & 2 &  &  &  \\ 
  UEMPLT5 & Civilians Unemployed: Less Than 5 Weeks & 5 &  &  &  \\ 
  UEMP5TO14 & Civilians Unemployed for 5-14 Weeks & 5 &  &  &  \\ 
  UEMP15OV & Civilians Unemployed: 15 Weeks \& Over & 5 &  &  &  \\ 
  UEMP15T26 & Civilians Unemployed for 15-26 Weeks & 5 &  &  &  \\ 
  UEMP27OV & Civilians Unemployed for 27 Weeks and Over & 5 &  &  &  \\ 
  PAYEMS & All Employees: Total nonfarm & 5 &  &  &  \\ 
  USGOOD & All Employees: Goods-Producing Industries & 5 &  &  &  \\ 
  CES1021000001 & All Employees: Mining and Logging: Mining & 5 &  &  &  \\ 
  USCONS & All Employees: Construction & 5 &  &  &  \\ 
  MANEMP & All Employees: Manufacturing & 5 &  &  &  \\ 
  DMANEMP & All Employees: Durable goods & 5 &  &  &  \\ 
  NDMANEMP & All Employees: Nondurable goods & 5 &  &  &  \\ 
  SRVPRD & All Employees: Service-Providing Industries & 5 &  &  &  \\ 
  USWTRADE & All Employees: Wholesale Trade & 5 &  &  &  \\ 
  USTRADE & All Employees: Retail Trade & 5 &  &  &  \\ 
  USFIRE & All Employees: Financial Activities & 5 &  &  &  \\ 
  USGOVT & All Employees: Government & 5 &  &  &  \\ 
  CES0600000007 & Avg Weekly Hours: Goods-Producing & 1 &  &  &  \\ 
  AWOTMAN & Avg Weekly Overtime Hourse: Manufacturing & 2 &  &  &  \\ 
  AWHMAN & Avg Weekly Hours: Manufacturing & 1 &  &  &  \\ 
  HOUST & Housing Starts: Total New Privately Owned & 4 &  &  &  \\ 
  HOUSTNE & Housing Starts, Northeast & 4 &  &  &  \\ 
  HOUSTMW & Housing Starts, Midwest & 4 &  &  &  \\ 
  HOUSTS & Housing Starts, South & 4 &  &  &  \\ 
  HOUSTW & Housing Starts, West & 4 &  &  &  \\ 
  PERMIT & New Private Housing Permits (SAAR) & 4 &  &  &  \\ 
  PERMITNE & New Private Housing Permits, Northeast (SAAR) & 4 &  &  &  \\ 
  PERMITMW & New Private Housing Permits, Midwest (SAAR) & 4 &  &  &  \\ 
  PERMITS & New Private Housing Permits, South (SAAR) & 4 &  &  &  \\ 
  PERMITW & New Private Housing Permits, West (SAAR & 4 &  &  &  \\ 
  AMDMNOx & New Orders for Durable goods & 5 &  &  &  \\ 
  ANDENOx & New Orders for Nondefense Capital goods & 5 &  &  &  \\ 
  AMDMUOx & Unfilled Orders for Durable goods & 5 &  &  &  \\ 
  BUSINVx & Total Business Inventories & 5 &  &  &  \\ 
  ISRATIOx & Total Business: Inventories to Sales Ratio & 2 &  &  &  \\ 
  M1SL & M1 Money Stock & 6 &  &  &  \\ 
  M2REAL & Real M2 Money Stock & 5 &  &  &  \\ 
  AMBSL & St. Louis Adjusted Monetary Base & 6 &  &  &  \\ 
  TOTRESNS & Total Reserves of Depository Institutions & 6 &  &  &  \\ 
  REALLN & Real Estate Loans at All Commerical Banks & 6 &  &  &  \\ 
  NONREVSL & Total Nonrevolving Credit & 6 &  &  &  \\ 
  S\&P: indust & S\&P's Common Stock Price Index: Industrials & 5 &  &  &  \\ 
  S\&P div yield & S\&P's Composite Common Stock: Dividend Yield & 2 &  & &  \\ 
  CP3Mx & 3-Month AA Financial Commercial Paper Rate & 2 &  &  &  \\ 
  TB3MS & 3-Month Treasury Bill & 2 &  &  &  \\ 
  TB6MS & 6-Month Treasury Bill & 2 &  &  &  \\ 
  GS1 & 1-Year Treasury Rate & 2 &  &  &  \\ 
  GS5 & 5-Year Treasury Rate & 2 &  &  &  \\ 
  AAA & Moody's Seasoned Aaa Corporate Bond Yield & 2 &  &  &  \\ 
  BAA & Moody's Seasoned Baa Corporate Bond Yield  & 2 &  &  &  \\ 
  COMPAPFFx & 3-Month Commercial Paper Minus FEDFUNDS & 1 &  &  &  \\ 
  TB3SMFFM & 3-Month Treasury C  Minus FEDFUNDS & 1 &  &  &  \\ 
  TB6SMFFM & 6-Month Treasury C  Minus FEDFUNDS & 1 &  &  &  \\ 
  T1YFFM & 1-Year Treasury C  Minus FEDFUNDS & 1 &  &  &  \\ 
  T5YFFM & 5-Year Treasury C  Minus FEDFUNDS & 1 &  &  &  \\ 
  AAAFFM & Moodys Aaa Corporate Bond  Minus FEDFUNDS & 1 &  &  &  \\ 
  BAAFFM & Moodys Baa Corporate Bond  Minus FEDFUNDS & 1 &  &  &  \\ 
  TWEXMMTH & Trade Weighted U.S. Dollar Index: Major Currencies & 5 &  &  &  \\ 
  EXSZUSx & Switzerland--U.S. Foreign Exchange Rate & 5 &  &  &  \\ 
  EXJPUSx & Japan--U.S. Foreign Exchange Rate & 5 &  &  &  \\ 
  EXCAUSx & Canada--U.S. Foreign Exchange Rate & 5 &  &  &  \\ 
  PPICMM & PPI: Metals and metal products & 6 &  &  &  \\ 
  CPIAPPSL & CPI: Apparel & 6 &  &  &  \\ 
  CPITRNSL & CPI: Transportation & 6 &  &  &  \\ 
  CPIMEDSL & CPI: Medical Care & 6 &  &  &  \\ 
  CUSR0000SAC & CPI: Commodities & 6 &  &  &  \\ 
  CUSR0000SAS & CPI: Services & 6 &  &  &  \\ 
  CPIULFSL & CPI: All Items Less Food & 6 &  &  &  \\ 
  CUSR0000SA0L5 & CPI: All Items Less Medical Care & 6 &  &  &  \\ 
  CES0600000008 & Avg Hourly Earnings: Goods-Producing & 6 &  &  &  \\ 
  CES2000000008 & Avg Hourly Earnings: Construction & 6 &  &  &  \\ 
  CES3000000008 & Avg Hourly Earnings: Manufacturing & 6 &  &  &  \\ 
  MZMSL & MZM Money Stock & 6 &  &  &  \\ 
  DTCOLNVHFNM & Consumer Motor Vehicle Loans Outstanding & 6 &  &  &  \\ 
  INVEST & Securities in Bank Credit at All Commercial Banks & 6 &  &  &  \\ 
\bottomrule
\end{longtable}
\end{tiny}

\clearpage
\subsection{EA-RTD data}
The Euro Area Real Time Database (EA-RTD) is maintained by the European Central Bank based on information from its \textit{Monthly Bulletin}, and available for download at \href{http://sdw.ecb.europa.eu/}{sdw.ecb.europa.eu}. A thorough description of the dataset is provided by \citet{giannone2012area}. All $165$ series are on a monthly frequency, with consistent coverage starting from 1999:01, while vintages are available from 2001:01. The dataset contains a substantial amount of variables in different units or transformations (e.g. unemployment numbers and rates), alongside series with substantial publication lags exceeding six months. We preselect the $94$ variables used for the forecasting application due to these reasons.

\begin{tiny}
\vspace*{0.5cm}
\begin{longtable}{p{2cm}p{6.5cm}p{0.5cm}C{1.2cm}C{1.2cm}C{1.2cm}}
\caption{Variables in the EA-RTD dataset.}\\[-1em]
\toprule
\textbf{Abbreviation} & \textbf{Description} & \textbf{Tc} & \textbf{Small} & \textbf{Medium} & \textbf{Large} \\ 
\midrule\endfirsthead
\textbf{Abbreviation} & \textbf{Description} & \textbf{Tc} & \textbf{Small} & \textbf{Medium} & \textbf{Large} \\ 
\midrule
\endhead

\midrule
\endfoot
\multicolumn{6}{p{15cm}}{\textit{Notes}: The dataset described in \citet{giannone2012area} is available for download at \href{http://sdw.ecb.europa.eu/}{sdw.ecb.europa.eu} (EA-RTD). Column \textbf{Tc} indicates the transformation of a series $x_t$ for obtaining stationarity: (1) no transformation, (2) $\Delta x_t$, (5) $\Delta \log(x_t)$, (6) $\Delta^2 \log(x_t)$ with $\Delta^i$ indicating $i$th differences. Columns Small, Medium and Large refer to different model sizes discussed in Section \ref{sec:data}.}
\endlastfoot
  EUR3M & Euribor 3-month & 2 & x & x & x \\ 
  C\_OV & HICP: Overall index & 5 & x & x & x\\  
  UNETO & Unemployment rate & 2 & x & x & x \\ 
  DJE50 & Dow Jones Euro Stoxx 50 Price Index & 5 &  & x & x \\ 
  XCONS & Industrial production & 5 &  & x & x \\ 
  10Y & Euro area 10-year Government Benchmark bond yield & 2 &  & x & x \\ 
  ERC0\_BGR & CPI deflated EER-38/Euro & 5 &  &  & x \\ 
  LOANSEC\_U\_NG & Total loans and securities & 6 &  &  & x \\ 
  M2\_V\_NC & Monetary aggregate M2 & 5 &  &  & x \\ 
  OILBR & Brent crude oil 1-month Forward & 6 &  &  & x \\ 
  10Y2Y & Spread between 10/2-year Government bond yield & 1 &  &  & x \\ 
  \midrule
  DJEBM & Dow Jones Euro Stoxx Basic Materials E Index & 5 &  &  &  \\ 
  DJECG & Dow Jones Euro Stoxx Consumer Goods Index & 5 &  &  &  \\ 
  DJEEN & Dow Jones Euro Stoxx Oil and Gas Energy Index & 5 &  &  &  \\ 
  DJEFI & Dow Jones Euro Stoxx Financials Index & 5 &  &  &  \\ 
  DJEHC & Dow Jones Euro Stoxx Healthcare Index & 5 &  &  &  \\ 
  DJEIG & Dow Jones Euro Stoxx Industrials Index & 5 &  &  &  \\ 
  DJENG & Dow Jones Euro Stoxx Consumer Services Index & 5 &  &  &  \\ 
  DJESB & Dow Jones Euro Stoxx Price Index & 5 &  &  &  \\ 
  DJETC & Dow Jones Euro Stoxx Telecommunications Index & 5 &  &  &  \\ 
  DJETE & Dow Jones Euro Stoxx Technology E Index & 5 &  &  &  \\ 
  DJEUT & Dow Jones Euro Stoxx Utilities E Index & 5 &  &  &  \\ 
  EONIA & Eonia rate & 2 &  &  &  \\ 
  EUR1M & Euribor 1-month & 2 &  &  &  \\ 
  EUR1Y & Euribor 1-year & 2 &  &  &  \\ 
  EUR6M & Euribor 6-month & 2 &  &  &  \\ 
  JPL3M & Japanese Yen 3-month BBBA Libor & 2 &  &  &  \\ 
  NIKKE & Nikkei 225 Stock Average Index & 5 &  &  &  \\ 
  SP500 & Standard and Poors 500 Composite Index & 5 &  &  &  \\ 
  USL3M & US Dollar 3-month BBA Libor & 2 &  &  &  \\ 
  EN00\_BGR & EER-38/Euro & 5 &  &  &  \\ 
  EN00\_NGR & EER-19/Euro & 5 &  &  &  \\ 
  ERC0\_NGR & CPI deflated EER-19/Euro & 5 &  &  &  \\ 
  ERP0\_NGR & PPI deflated EER-19/Euro & 5 &  &  &  \\ 
  LOAN\_U\_NG & Loans excluding reverse repos & 6 &  &  &  \\ 
  M1\_V\_NC & Monetary aggregate M1 & 5 &  &  &  \\ 
  M3\_V\_NC & Monetary aggregate M3 & 5 &  &  &  \\ 
  C\_COMM & HICP: Communication services & 6 &  &  &  \\ 
  C\_FOOD & HICP: Food incl. alcohol and tobacco & 6 &  &  &  \\ 
  C\_FOODPR & HICP: Processed food incl. alcohol and tobacco & 6 &  &  &  \\ 
  C\_FOODUN & HICP: Unprocessed food & 6 &  &  &  \\ 
  C\_GOOD & HICP: Goods & 6 &  &  &  \\ 
  C\_HOUS & HICP: Housing services & 6 &  &  &  \\ 
  C\_IGXE & HICP: Industrial goods excluding energy & 6 &  &  &  \\ 
  C\_MISC & HICP: Miscellaneous services & 6 &  &  &  \\ 
  C\_NRGY & HICP: Energy & 6 &  &  &  \\ 
  C\_XEFUN & HICP: All-items excluding energy and unprocessed food & 6 &  &  &  \\
  C\_RECR & HICP: Recreation and personal services & 6 &  &  &  \\ 
  C\_SERV & HICP: Services & 6 &  &  &  \\ 
  C\_TRAN & HICP: Transport services & 6 &  &  &  \\ 
  P\_CAPGO\_DS & Producer price index, MIG Capital Goods Industry & 6 &  &  &  \\ 
  P\_CONGO\_DS & Producer price index, Consumer goods industry & 6 &  &  &  \\ 
  P\_INTGO\_DS & Producer price index, MIG Intermediate Goods & 6 &  &  &  \\ 
  P\_MANUF\_DS & Producer price index, Manufacturing & 6 &  &  &  \\ 
  P\_XCONS\_DS & Industrial producer prices (excl. construction) & 6 &  &  &  \\ 
  FETOT & Unemployment rate: Total female & 2 &  &  &  \\ 
  FMO25 & Unemployment rate: Age 25 and over male/female & 2 &  &  &  \\ 
  FMU25 & Unemployment rate: Age und 25 male/female & 2 &  &  &  \\ 
  MATOT & Unemployment rate: Total male & 2 &  &  &  \\ 
  COAOB & Assessment of order books & 1 &  &  &  \\ 
  COCCI & Construction Confidence Indicator & 1 &  &  &  \\ 
  COEEX & Employment expectations for the months ahead & 1 &  &  &  \\ 
  CSCCI & Consumer Confidence Indicator & 1 &  &  &  \\ 
  CSF12 & Financial situation over next 12 months & 1 &  &  &  \\ 
  CSG12 & General economic situation of next 12 months & 1 &  &  &  \\ 
  CSS12 & Savings over next 12 months & 1 &  &  &  \\ 
  CSU12 & Unemployment expectations over next 12 months & 1 &  &  &  \\ 
  ESIND & Economic Sentiment Indicator & 1 &  &  &  \\ 
  ISAOB & Assessment of order-book levels & 1 &  &  &  \\ 
  ISASP & Assessment of stocks of finished products & 1 &  &  &  \\ 
  ISICI & Industrial Confidence Indicator & 1 &  &  &  \\ 
  ISPEX & Production expectations for the months ahead & 1 &  &  &  \\ 
  RSAOS & Assessment of stocks & 1 &  &  &  \\ 
  RSBEX & Expected business situation & 1 &  &  &  \\ 
  RSPBS & Present business situation & 1 &  &  &  \\ 
  RSRCI & Retail Confidence Indicator & 1 &  &  &  \\ 
  SSABC & Assessment of the business climate & 1 &  &  &  \\ 
  SSEDE & Evolution of demand expected in the months ahead & 1 &  &  &  \\ 
  SSEDR & Evolution of demand in recent months & 1 &  &  &  \\ 
  SSSCI & Sevice Confidence Indicator & 1 &  &  &  \\ 
  CAPGO & Industrial production: capital goods & 5 &  &  &  \\ 
  CONGO & Industrial production: consumer goods & 5 &  &  &  \\ 
  CONST & Construction production & 5 &  &  &  \\ 
  DCOGO & Industrial production: durable consumer goods & 5 &  &  &  \\ 
  ENERG & Industrial production: energy & 5 &  &  &  \\ 
  INDTO & Industrial production: total including construction & 5 &  &  &  \\ 
  INTGO & Industrial production: intermediate goods & 5 &  &  &  \\ 
  MANUF & Industrial production: manufacturing & 5 &  &  &  \\ 
  NCOGO & Industrial production: non-durable consumer goods & 5 &  &  &  \\ 
  XCOEN & Industrial production: total excl. construction, energy & 5 &  &  &  \\ 
  NCARS & New passenger car registration & 5 &  &  &  \\ 
  RTFOBETD & Retail trade turnover: food, beverages and tobacco & 5 &  &  &  \\ 
  RTHOUSED & Retail trade turnover: audio and video equipment etc. & 5 &  &  &  \\ 
  RTNONFOD & Retail trade turnover: non-food products & 5 &  &  &  \\ 
\bottomrule
\end{longtable}
\end{tiny}

\setcounter{equation}{0}
\renewcommand\theequation{C.\arabic{equation}}
\section{Additional results}\label{app:C}
This appendix provides additional information for variable specific forecast performance measures to be considered in the context of Sections \ref{sec:forecasts_us} and \ref{sec:forecasts_ea}. 

\begin{figure}[!htbp]
\begin{subfigure}[b]{0.48\textwidth}
\caption{Point forecasts}
\includegraphics[width=\textwidth]{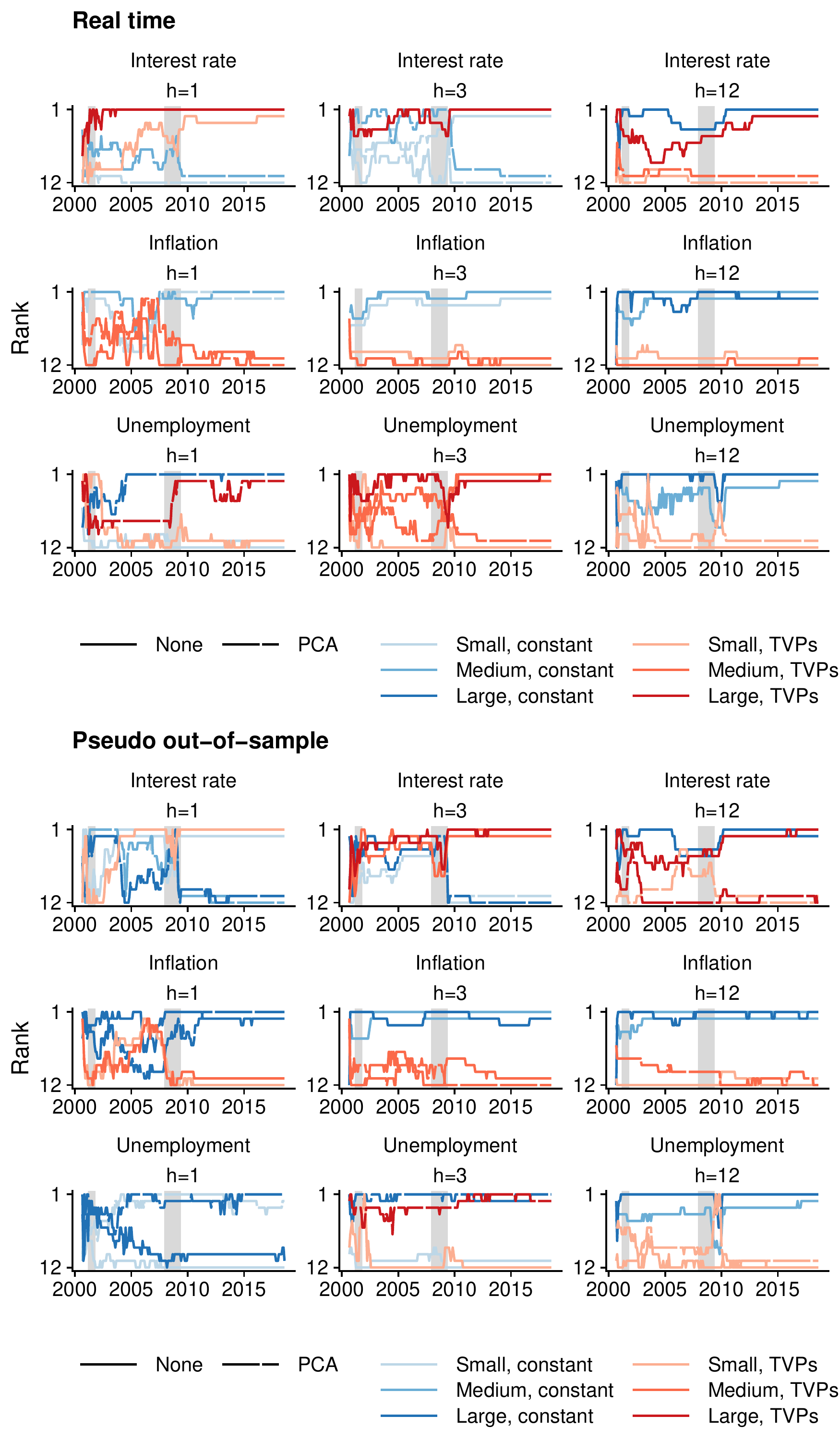}
\end{subfigure}
\begin{subfigure}[b]{0.48\textwidth}
\caption{Density forecasts}
\includegraphics[width=\textwidth]{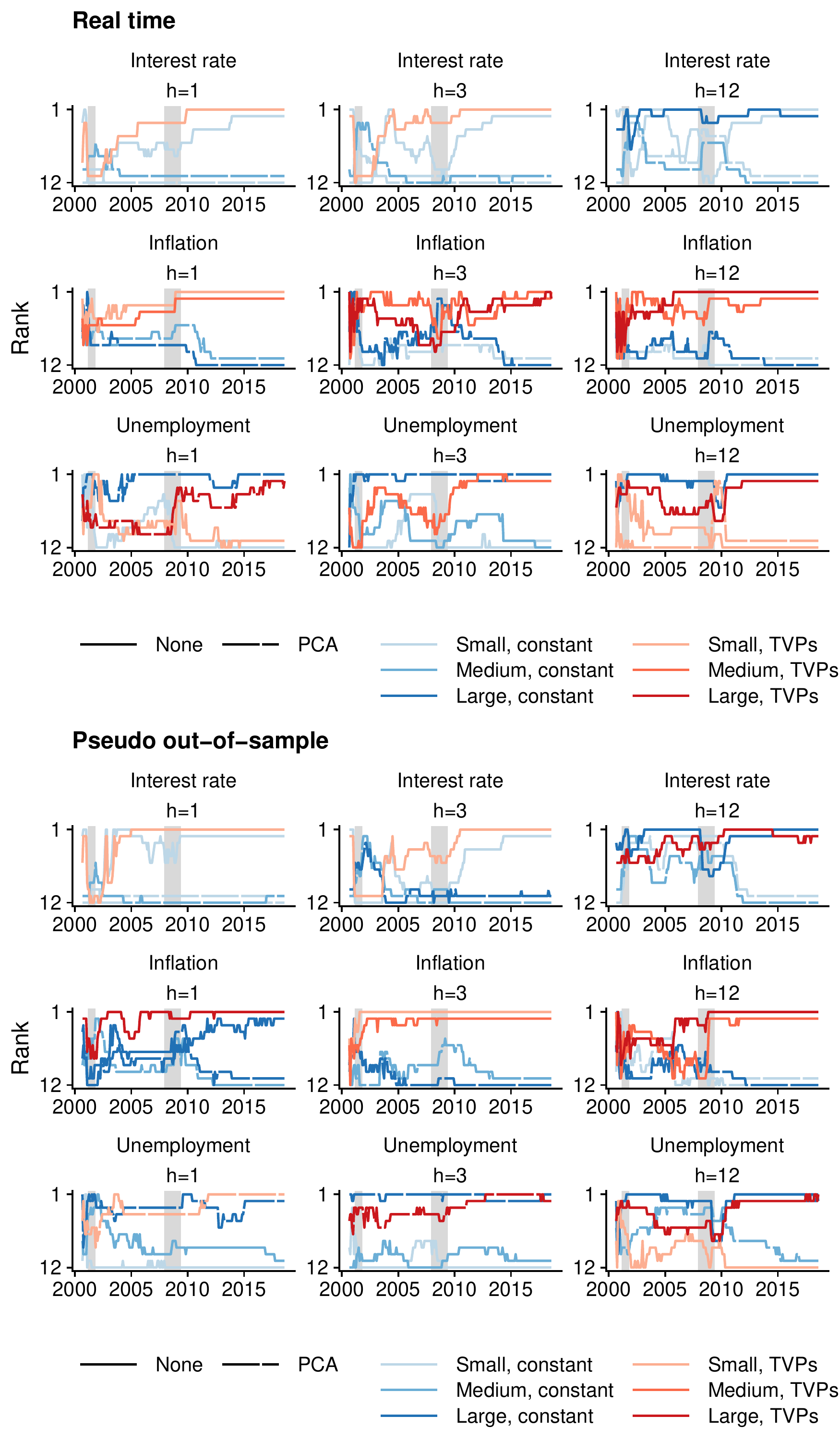}
\end{subfigure}
\caption{Performance ranking for point and density forecast for the US.}\label{fig:app_USranks}\vspace*{-0.3cm}
\caption*{\footnotesize\textit{Note}: Ranks are derived from cumulative absolute forecast errors (point forecasts) and cumulative marginal log predictive scores (density forecasts). The figures show the respective two best and worst performing models. The grey shaded areas indicate recessions dated by the NBER Business Cycle Dating Committee.}
\end{figure}

\begin{figure}[!htbp]
\begin{subfigure}[b]{0.48\textwidth}
\caption{Point forecasts}
\includegraphics[width=\textwidth]{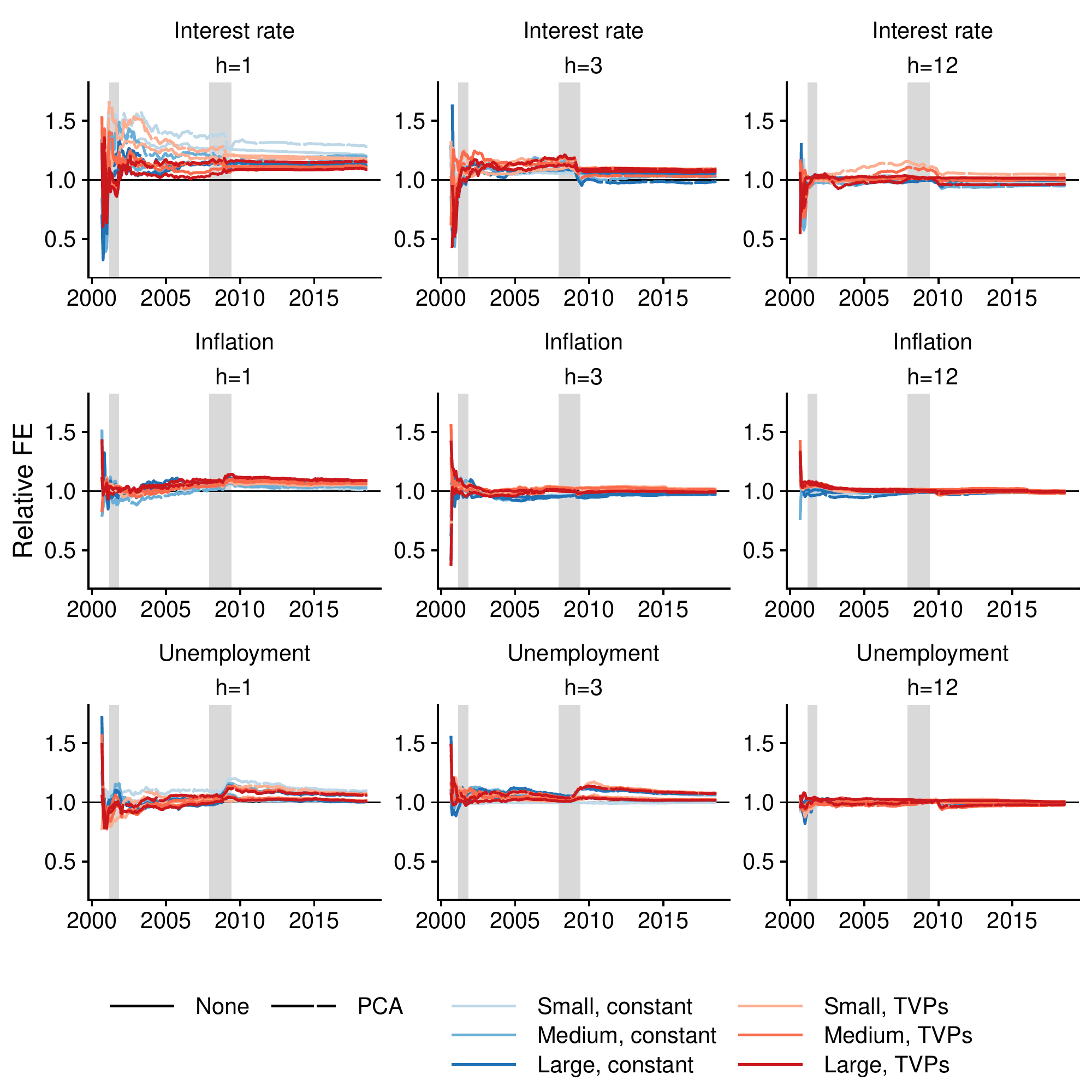}
\end{subfigure}
\begin{subfigure}[b]{0.48\textwidth}
\caption{Density forecasts}
\includegraphics[width=\textwidth]{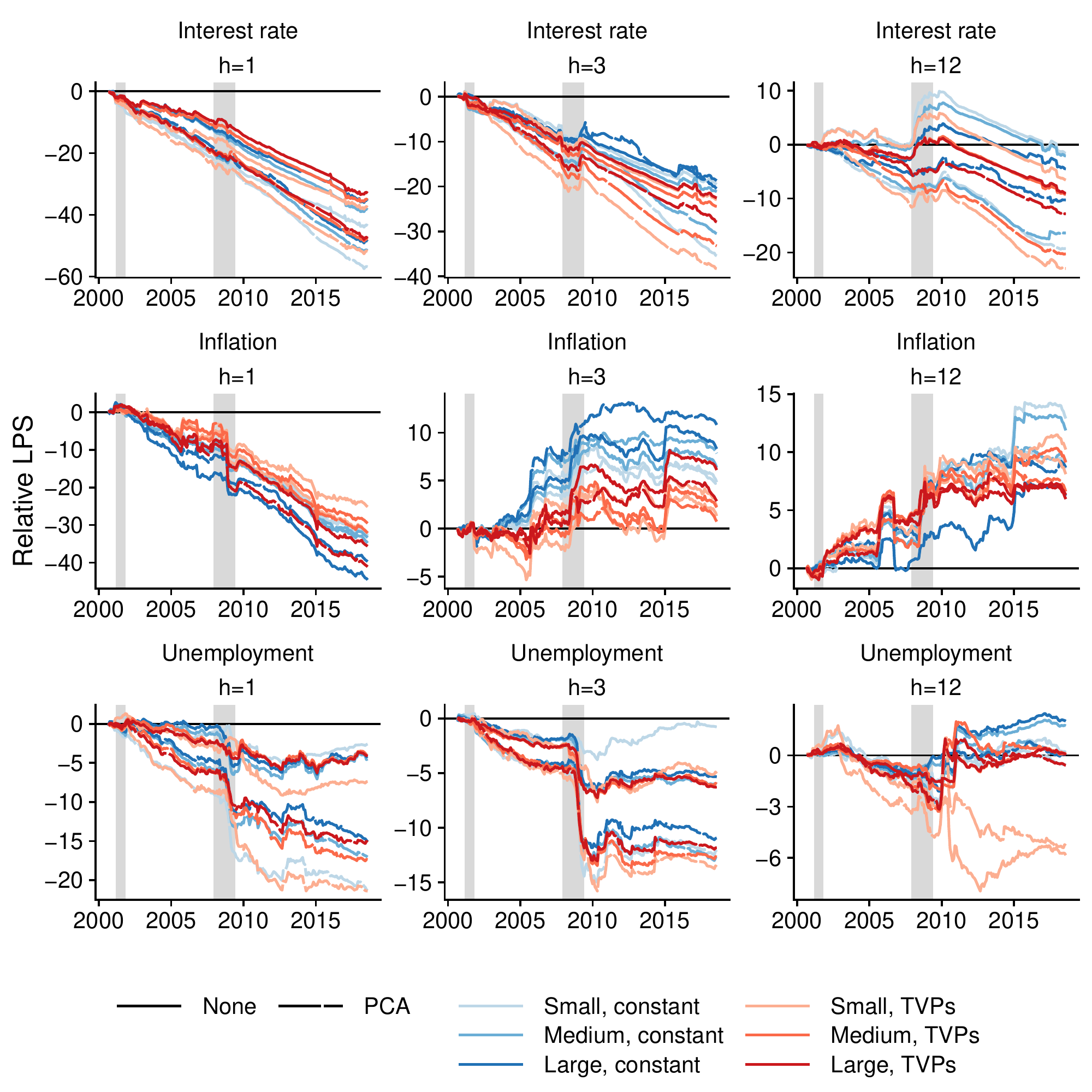}
\end{subfigure}
\caption{Relative performance measures for real time and pseudo out-of-sample information sets for the US.}\label{fig:app_USdiffs}\vspace*{-0.3cm}
\caption*{\footnotesize\textit{Note}: Each real time model is benchmarked against its complementary specification estimated using the pseudo out-of-sample information set (differences for density forecasts, ratios for point forecasts). The grey shaded areas indicate recessions dated by the NBER Business Cycle Dating Committee.}
\end{figure}

\begin{figure}[!htbp]
\begin{subfigure}[b]{0.48\textwidth}
\caption{Point forecasts}
\includegraphics[width=\textwidth]{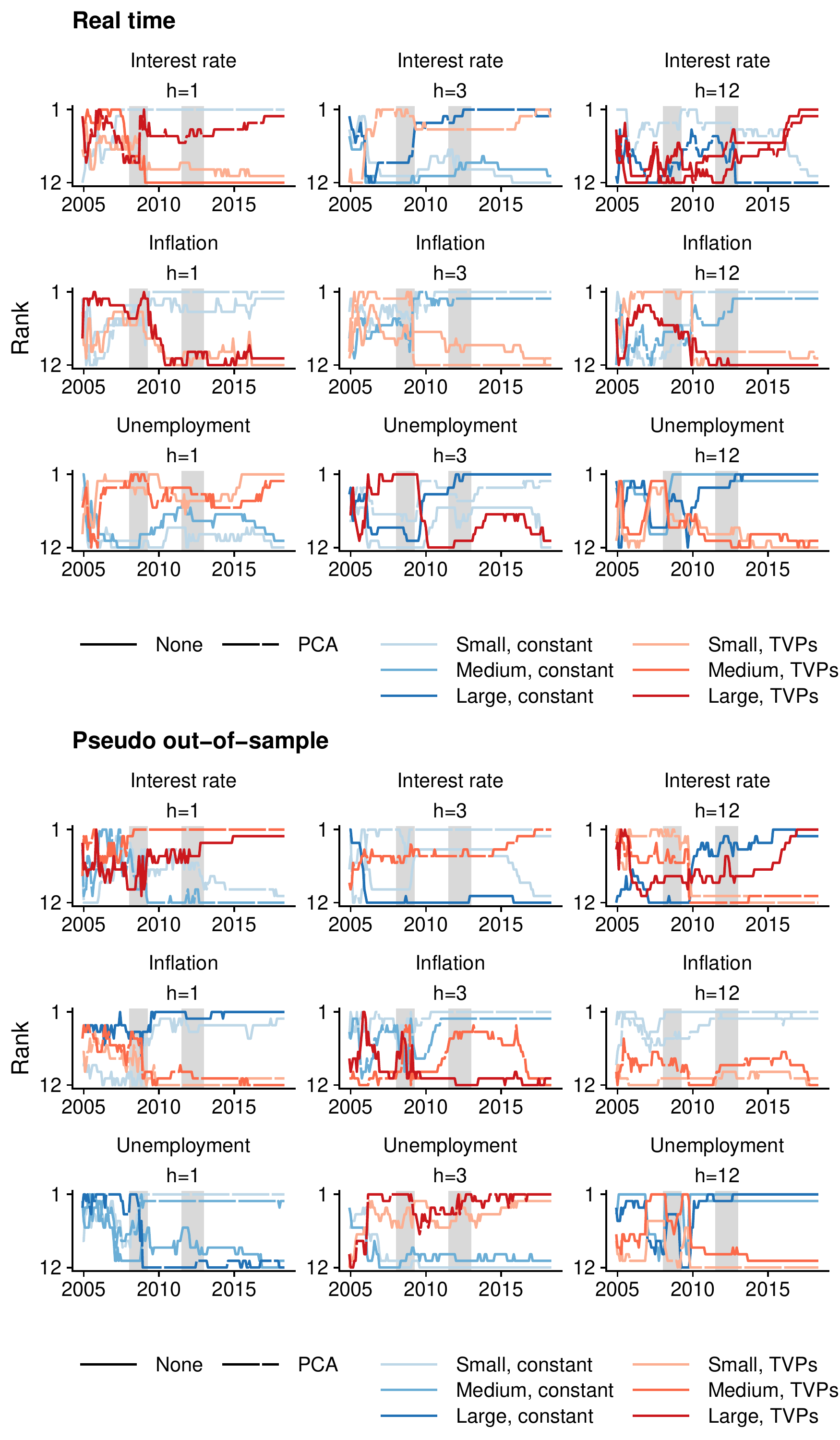}
\end{subfigure}
\begin{subfigure}[b]{0.48\textwidth}
\caption{Density forecasts}
\includegraphics[width=\textwidth]{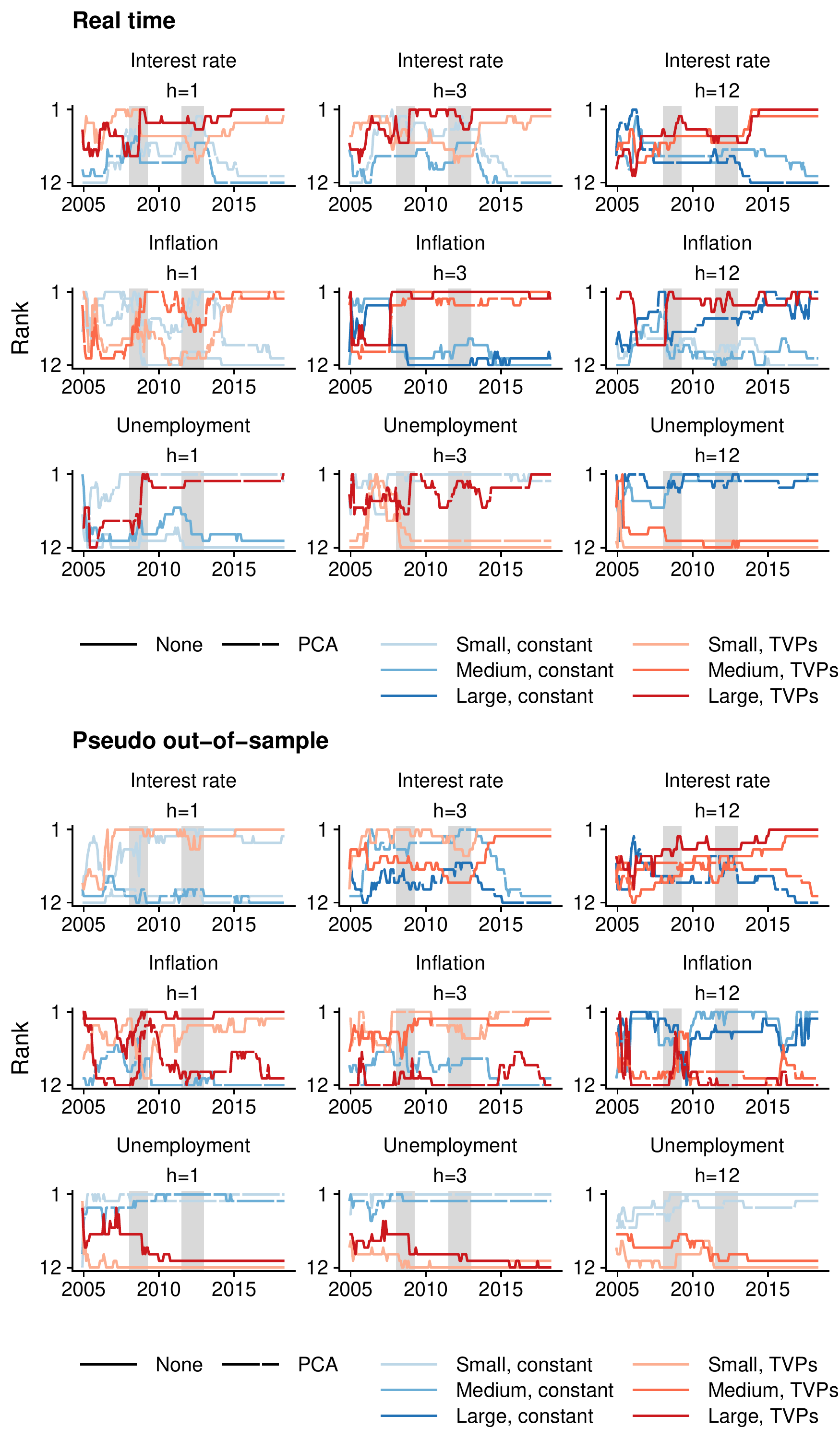}
\end{subfigure}
\caption{Performance ranking for point and density forecast for the EA.}\label{fig:app_EAranks}\vspace*{-0.3cm}
\caption*{\footnotesize\textit{Note}: Ranks are derived from cumulative absolute forecast errors (point forecasts) and cumulative marginal log predictive scores (density forecasts). The figures show the respective two best and worst performing models. The grey shaded areas indicate recessions dated by the CEPR Euro Area Business Cycle Dating Committee.}
\end{figure}

\begin{figure}[!htbp]
\begin{subfigure}[b]{0.48\textwidth}
\caption{Point forecasts}
\includegraphics[width=\textwidth]{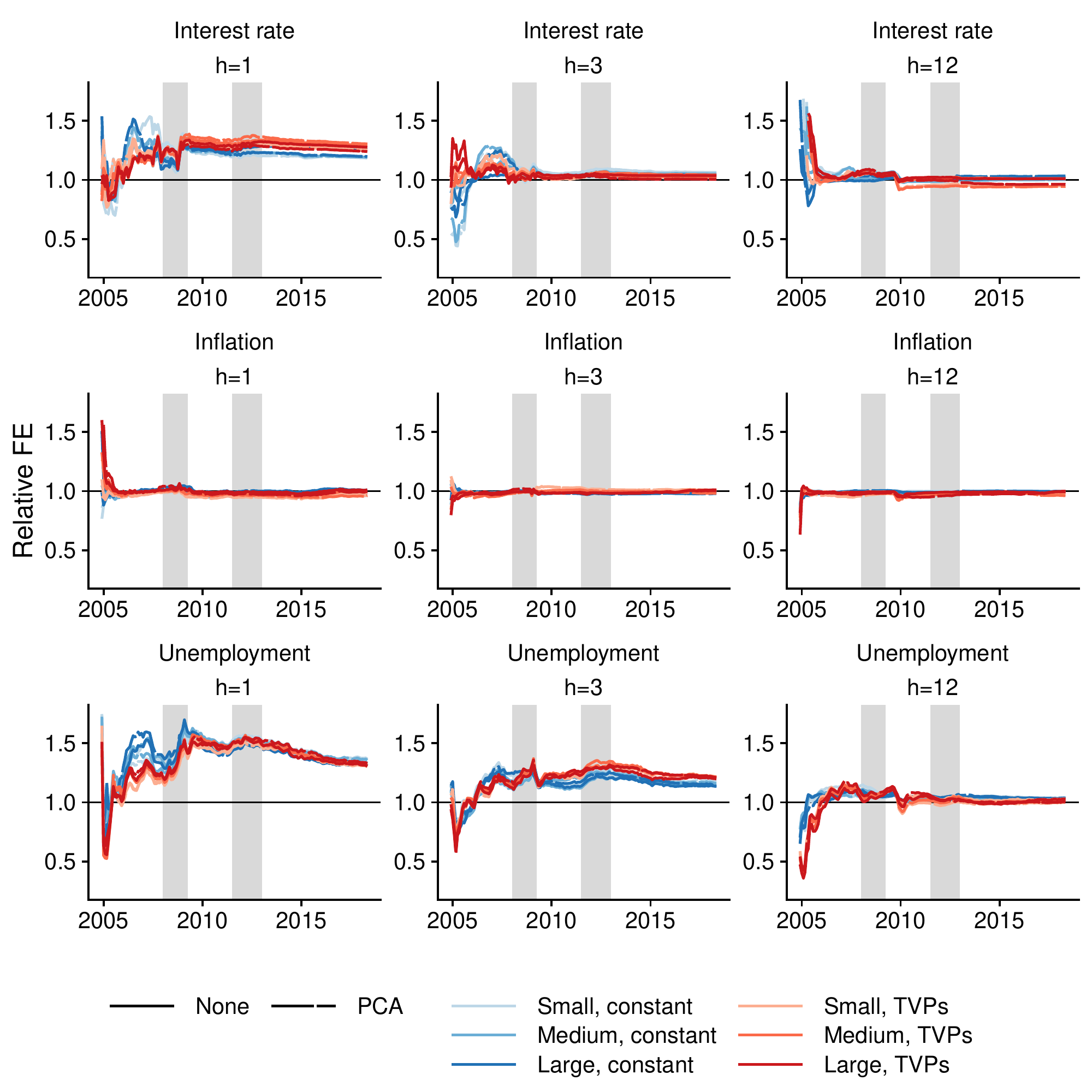}
\end{subfigure}
\begin{subfigure}[b]{0.48\textwidth}
\caption{Density forecasts}
\includegraphics[width=\textwidth]{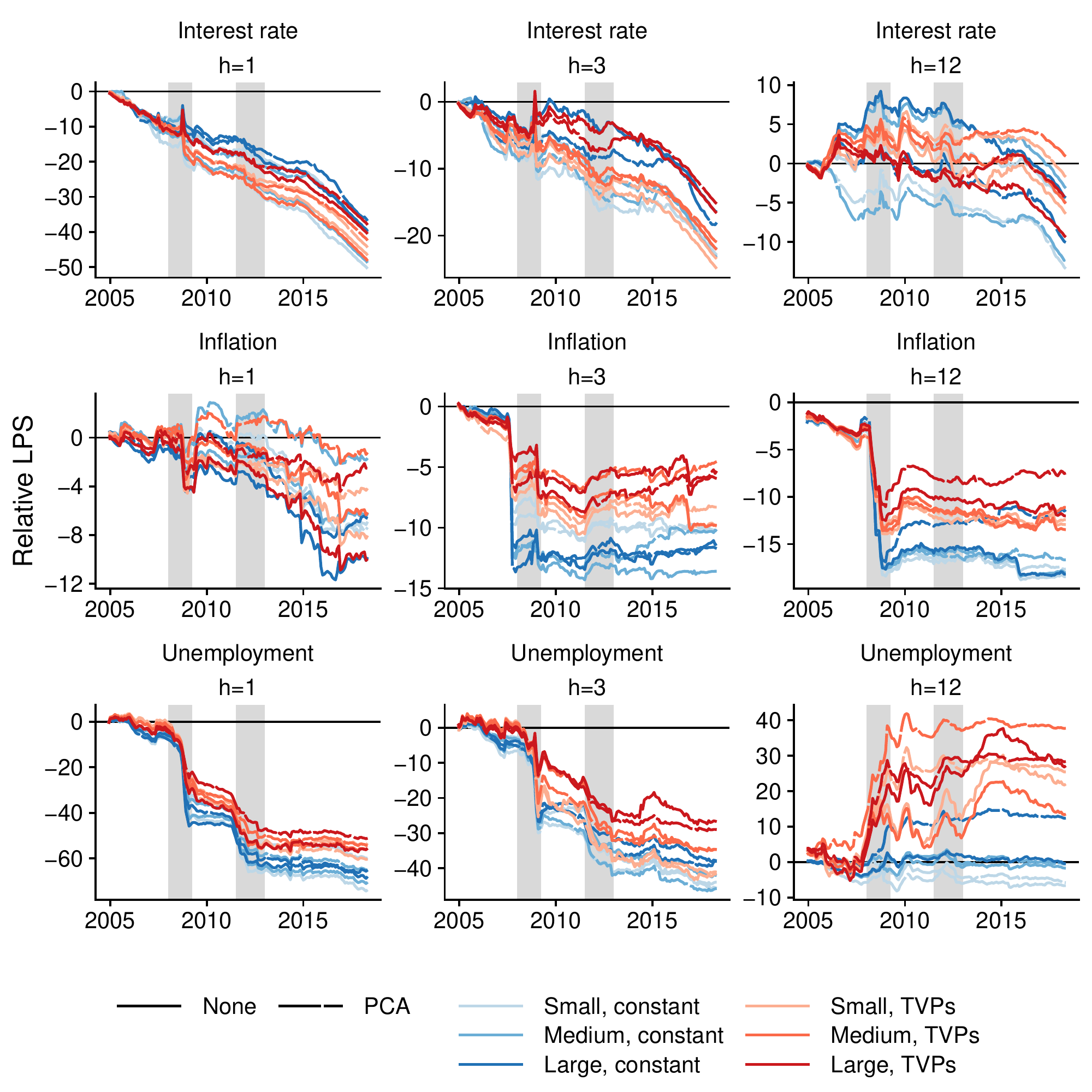}
\end{subfigure}
\caption{Relative performance measures for real time and pseudo out-of-sample information sets for the EA.}\label{fig:app_EAdiffs}\vspace*{-0.3cm}
\caption*{\footnotesize\textit{Note}: Each real time model is benchmarked against its complementary specification estimated using the pseudo out-of-sample information set (differences for density forecasts, ratios for point forecasts). The grey shaded areas indicate recessions dated by the CEPR Euro Area Business Cycle Dating Committee.}
\end{figure}

Figures \ref{fig:app_USranks} (US) and \ref{fig:app_EAranks} (EA) show the model ranks for point and density forecasts for real time and pseudo out-of-sample simulations over time. In Figs. \ref{fig:app_USdiffs} (US) and \ref{fig:app_EAdiffs} (EA), each real time model is benchmarked against its complementary specification estimated using the pseudo out-of-sample information set. The relative differences and ratios can be interpreted as measures capturing the distance between real time and pseudo out-of-sample forecasts, and do not necessarily indicate superior forecast performance, since each real time model is benchmarked against its pseudo out-of-sample counterpart. 

Corresponding average measures over the full holdout are displayed in Tabs. \ref{tab:fcst_us} (US) and \ref{tab:fcst_ea} (EA). The tables feature average LPSs and RMSEs for the real time information set with the rank at the end of the holdout in parentheses in the first row per model. The second row indicates the difference (LPSs) and ratio (RMSEs) between real time and pseudo out-of-sample simulations within each model specification, with ranks for pseudo out-of-sample exercises in parentheses.

\begin{table*}[ht]
\begin{center}
\caption{Average marginal LPSs and RMSEs for the US.}\label{tab:fcst_us}\vspace*{-0.75em}
\begin{threeparttable}
\tiny
\begin{tabular*}{\textwidth}{@{\extracolsep{\fill}} ld{1.3}d{1.3}d{1.3}d{1.3}d{1.3}d{1.3}}
\toprule
                        & \multicolumn{3}{c}{LPS} & \multicolumn{3}{c}{RMSE}\\
                        \cmidrule(lr){2-4}\cmidrule(lr){5-7}
\textbf{Model/Variable} & \multicolumn{1}{c}{$h=1$} & \multicolumn{1}{c}{$h=3$} & \multicolumn{1}{c}{$h=12$} & \multicolumn{1}{c}{$h=1$} & \multicolumn{1}{c}{$h=3$} & \multicolumn{1}{c}{$h=12$} \\ 
  \midrule
\textbf{Interest rate} & & & & & & \\
Small, CPs & 0.986 (2) & 0.885 (2) & 0.754 (1) & 0.147 (10) & 0.163 (11) & 0.168 (6) \\ 
   & -0.202 (2) & -0.090 (2) & -0.007 (3) & 1.203 (8) & 1.037 (9) & 1.001 (7) \\ 
  Medium, CPs & 0.951 (5) & 0.841 (6) & 0.701 (5) & 0.137 (4) & 0.158 (9) & 0.169 (7) \\ 
   & -0.178 (4) & -0.098 (5) & -0.010 (6) & 1.147 (2) & 1.068 (4) & 1.005 (6) \\ 
  Large, CPs & 0.946 (6) & 0.850 (5) & 0.743 (2) & 0.137 (5) & 0.158 (8) & 0.161 (3) \\ 
   & -0.164 (6) & -0.086 (6) & -0.021 (1) & 1.142 (5) & 1.065 (3) & 1.001 (3) \\ 
  Small, PCA, CPs & 0.447 (12) & 0.398 (12) & 0.302 (11) & 0.163 (12) & 0.167 (12) & 0.165 (5) \\ 
   & -0.263 (12) & -0.164 (10) & -0.089 (11) & 1.204 (12) & 1.011 (11) & 0.998 (5) \\ 
  Medium, PCA, CPs & 0.478 (11) & 0.409 (11) & 0.300 (12) & 0.148 (11) & 0.157 (7) & 0.160 (2) \\ 
   & -0.239 (11) & -0.141 (12) & -0.076 (12) & 1.128 (10) & 0.954 (10) & 0.979 (4) \\ 
  Large, PCA, CPs & 0.551 (10) & 0.467 (10) & 0.380 (10) & 0.146 (9) & 0.157 (6) & 0.154 (1) \\ 
   & -0.225 (10) & -0.095 (11) & -0.048 (10) & 1.100 (11) & 0.928 (12) & 0.990 (1) \\ 
  Small, TVPs & 1.016 (1) & 0.902 (1) & 0.717 (4) & 0.143 (8) & 0.160 (10) & 0.172 (10) \\ 
   & -0.174 (1) & -0.106 (1) & -0.030 (4) & 1.197 (1) & 1.052 (6) & 1.006 (10) \\ 
  Medium, TVPs & 0.976 (3) & 0.859 (4) & 0.676 (6) & 0.135 (2) & 0.156 (5) & 0.171 (9) \\ 
   & -0.163 (3) & -0.113 (3) & -0.043 (5) & 1.126 (3) & 1.083 (1) & 1.012 (8) \\ 
  Large, TVPs & 0.972 (4) & 0.863 (3) & 0.721 (3) & 0.136 (3) & 0.156 (4) & 0.162 (4) \\ 
   & -0.153 (5) & -0.104 (4) & -0.042 (2) & 1.115 (7) & 1.066 (2) & 1.011 (2) \\ 
  Small, PCA, TVPs & 0.792 (8) & 0.692 (8) & 0.483 (8) & 0.141 (7) & 0.154 (3) & 0.180 (12) \\ 
   & -0.240 (8) & -0.177 (7) & -0.106 (8) & 1.165 (6) & 1.033 (5) & 1.023 (12) \\ 
  Medium, PCA, TVPs & 0.788 (9) & 0.675 (9) & 0.465 (9) & 0.133 (1) & 0.148 (1) & 0.174 (11) \\ 
   & -0.222 (9) & -0.154 (9) & -0.094 (9) & 1.043 (9) & 0.963 (8) & 1.013 (11) \\ 
  Large, PCA, TVPs & 0.827 (7) & 0.732 (7) & 0.571 (7) & 0.138 (6) & 0.151 (2) & 0.171 (8) \\ 
   & -0.220 (7) & -0.129 (8) & -0.059 (7) & 1.144 (4) & 0.989 (7) & 1.003 (9) \\ 
  \midrule
  \textbf{Inflation} & & & & & & \\
  Small, CPs & -0.149 (7) & -0.143 (6) & -0.141 (6) & 0.337 (4) & 0.323 (3) & 0.321 (3) \\ 
   & -0.153 (8) &  0.022 (7) &  0.060 (8) & 1.075 (6) & 0.980 (3) & 0.998 (3) \\ 
  Medium, CPs & -0.157 (9) & -0.151 (8) & -0.156 (9) & 0.336 (2) & 0.321 (1) & 0.318 (1) \\ 
   & -0.161 (9) &  0.029 (9) &  0.055 (9) & 1.083 (4) & 0.980 (1) & 0.997 (1) \\ 
  Large, CPs & -0.151 (8) & -0.131 (4) & -0.153 (8) & 0.341 (5) & 0.322 (2) & 0.321 (2) \\ 
   & -0.206 (2) &  0.038 (8) &  0.029 (7) & 1.122 (2) & 0.981 (2) & 1.004 (2) \\ 
  Small, PCA, CPs & -0.167 (10) & -0.175 (11) & -0.196 (11) & 0.332 (1) & 0.327 (4) & 0.323 (4) \\ 
   & -0.151 (10) &  0.021 (10) &  0.040 (11) & 1.062 (5) & 0.985 (4) & 0.991 (4) \\ 
  Medium, PCA, CPs & -0.188 (11) & -0.172 (10) & -0.192 (10) & 0.337 (3) & 0.328 (5) & 0.324 (6) \\ 
   & -0.154 (12) &  0.036 (11) &  0.042 (10) & 1.087 (3) & 0.980 (5) & 0.993 (6) \\ 
  Large, PCA, CPs & -0.213 (12) & -0.185 (12) & -0.211 (12) & 0.343 (6) & 0.328 (6) & 0.324 (5) \\ 
   & -0.184 (11) &  0.050 (12) &  0.040 (12) & 1.138 (1) & 0.970 (6) & 0.994 (5) \\ 
  Small, TVPs & -0.077 (1) & -0.107 (3) & -0.102 (3) & 0.359 (10) & 0.366 (12) & 0.358 (12) \\ 
   & -0.116 (5) &  0.006 (1) &  0.047 (3) & 1.065 (12) & 1.011 (10) & 1.004 (12) \\ 
  Medium, TVPs & -0.083 (2) & -0.106 (1) & -0.101 (2) & 0.360 (11) & 0.365 (11) & 0.358 (11) \\ 
   & -0.137 (3) &  0.010 (2) &  0.042 (2) & 1.072 (11) & 1.004 (11) & 1.005 (10) \\ 
  Large, TVPs & -0.099 (3) & -0.106 (2) & -0.093 (1) & 0.358 (8) & 0.353 (7) & 0.346 (8) \\ 
   & -0.165 (1) &  0.028 (3) &  0.029 (1) & 1.101 (8) & 0.985 (7) & 1.000 (7) \\ 
  Small, PCA, TVPs & -0.117 (4) & -0.148 (7) & -0.141 (5) & 0.355 (7) & 0.358 (9) & 0.347 (9) \\ 
   & -0.149 (6) &  0.011 (5) &  0.038 (6) & 1.070 (9) & 0.993 (9) & 0.983 (9) \\ 
  Medium, PCA, TVPs & -0.119 (5) & -0.140 (5) & -0.139 (4) & 0.364 (12) & 0.365 (10) & 0.350 (10) \\ 
   & -0.149 (7) &  0.003 (4) &  0.031 (4) & 1.097 (10) & 0.998 (12) & 0.982 (11) \\ 
  Large, PCA, TVPs & -0.137 (6) & -0.151 (9) & -0.143 (7) & 0.358 (9) & 0.353 (8) & 0.344 (7) \\ 
   & -0.190 (4) &  0.013 (6) &  0.028 (5) & 1.124 (7) & 0.986 (8) & 0.988 (8) \\ 
  \midrule
  \textbf{Unemployment} & & & & & & \\
  Small, CPs & 0.442 (12) & 0.430 (11) & 0.387 (4) & 0.161 (12) & 0.163 (12) & 0.163 (5) \\ 
   & -0.012 (12) & -0.004 (12) &  0.001 (4) & 0.997 (12) & 0.999 (12) & 0.996 (6) \\ 
  Medium, CPs & 0.452 (8) & 0.426 (12) & 0.380 (6) & 0.155 (11) & 0.160 (10) & 0.163 (6) \\ 
   & -0.021 (11) & -0.028 (11) &  0.008 (6) & 1.019 (10) & 1.020 (11) & 1.002 (4) \\ 
  Large, CPs & 0.455 (7) & 0.431 (10) & 0.405 (1) & 0.155 (10) & 0.161 (11) & 0.160 (1) \\ 
   & -0.019 (10) & -0.025 (10) &  0.009 (2) & 1.022 (9) & 1.026 (10) & 0.997 (1) \\ 
  Small, PCA, CPs & 0.446 (10) & 0.440 (9) & 0.363 (9) & 0.155 (9) & 0.156 (7) & 0.167 (8) \\ 
   & -0.099 (3) & -0.057 (6) & -0.003 (9) & 1.112 (3) & 1.056 (6) & 1.002 (7) \\ 
  Medium, PCA, CPs & 0.465 (5) & 0.446 (5) & 0.361 (10) & 0.151 (4) & 0.156 (4) & 0.167 (7) \\ 
   & -0.079 (4) & -0.059 (4) &  0.001 (11) & 1.079 (4) & 1.066 (5) & 1.000 (8) \\ 
  Large, PCA, CPs & 0.479 (1) & 0.459 (1) & 0.385 (5) & 0.148 (1) & 0.154 (1) & 0.163 (4) \\ 
   & -0.070 (2) & -0.051 (1) &  0.000 (5) & 1.067 (2) & 1.057 (2) & 0.997 (5) \\ 
  Small, TVPs & 0.444 (11) & 0.443 (8) & 0.313 (12) & 0.153 (8) & 0.158 (9) & 0.180 (12) \\ 
   & -0.034 (9) & -0.023 (9) & -0.027 (12) & 1.008 (11) & 1.031 (9) & 1.029 (12) \\ 
  Medium, TVPs & 0.469 (3) & 0.455 (2) & 0.391 (3) & 0.151 (3) & 0.154 (2) & 0.161 (3) \\ 
   & -0.018 (8) & -0.028 (7) &  0.000 (3) & 1.006 (8) & 1.030 (7) & 0.998 (3) \\ 
  Large, TVPs & 0.469 (4) & 0.447 (4) & 0.396 (2) & 0.151 (5) & 0.156 (6) & 0.161 (2) \\ 
   & -0.019 (7) & -0.029 (8) &  0.000 (1) & 1.017 (7) & 1.025 (8) & 1.001 (2) \\ 
  Small, PCA, TVPs & 0.451 (9) & 0.444 (7) & 0.342 (11) & 0.152 (7) & 0.156 (5) & 0.176 (11) \\ 
   & -0.099 (1) & -0.063 (3) & -0.024 (8) & 1.102 (1) & 1.072 (3) & 1.009 (11) \\ 
  Medium, PCA, TVPs & 0.460 (6) & 0.444 (6) & 0.363 (8) & 0.152 (6) & 0.158 (8) & 0.168 (9) \\ 
   & -0.080 (5) & -0.059 (5) &  0.000 (10) & 1.083 (6) & 1.082 (4) & 0.980 (10) \\ 
  Large, PCA, TVPs & 0.470 (2) & 0.454 (3) & 0.365 (7) & 0.149 (2) & 0.154 (3) & 0.168 (10) \\ 
   & -0.070 (6) & -0.055 (2) & -0.003 (7) & 1.069 (5) & 1.068 (1) & 0.983 (9) \\ 
   \bottomrule
\end{tabular*}
\begin{tablenotes}[para,flushleft]
\tiny{\textit{Notes}: The first row per model shows average LPSs and RMSEs for the real time information set with the rank at the end of the holdout in parentheses. The second row indicates the difference (LPSs) and ratio (RMSEs) between real time and pseudo out-of-sample simulations within each model specification, with ranks for pseudo out-of-sample exercises in parentheses.}
\end{tablenotes}
\end{threeparttable}
\end{center}
\end{table*}

\begin{table*}[ht]
\begin{center}
\caption{Average marginal LPSs and RMSEs for the EA.}\label{tab:fcst_ea}\vspace*{-0.75em}
\begin{threeparttable}
\tiny
\begin{tabular*}{\textwidth}{@{\extracolsep{\fill}} ld{1.3}d{1.3}d{1.3}d{1.3}d{1.3}d{1.3}}
\toprule
                        & \multicolumn{3}{c}{LPS} & \multicolumn{3}{c}{RMSE}\\
                        \cmidrule(lr){2-4}\cmidrule(lr){5-7}
\textbf{Model/Variable} & \multicolumn{1}{c}{$h=1$} & \multicolumn{1}{c}{$h=3$} & \multicolumn{1}{c}{$h=12$} & \multicolumn{1}{c}{$h=1$} & \multicolumn{1}{c}{$h=3$} & \multicolumn{1}{c}{$h=12$} \\ 
  \midrule
\textbf{Interest rate} & & & & & & \\
Small, CPs & 1.405 (6) & 1.183 (7) & 0.918 (9) & 0.135 (7) & 0.150 (8) & 0.160 (3) \\ 
   & -0.312 (2) & -0.141 (6) & -0.031 (9) & 1.290 (2) & 1.088 (3) & 1.011 (3) \\ 
  Medium, CPs & 1.404 (7) & 1.194 (6) & 0.947 (6) & 0.133 (4) & 0.148 (5) & 0.159 (1) \\ 
   & -0.301 (4) & -0.136 (4) & -0.019 (8) & 1.255 (4) & 1.069 (4) & 1.010 (1) \\ 
  Large, CPs & 1.436 (3) & 1.219 (5) & 0.952 (5) & 0.136 (8) & 0.151 (11) & 0.161 (6) \\ 
   & -0.246 (6) & -0.102 (8) & -0.027 (6) & 1.246 (8) & 1.046 (8) & 1.017 (5) \\ 
  Small, PCA, CPs & 1.337 (11) & 1.135 (11) & 0.901 (10) & 0.133 (3) & 0.146 (3) & 0.162 (8) \\ 
   & -0.248 (11) & -0.142 (10) & -0.083 (5) & 1.286 (1) & 1.093 (1) & 1.023 (6) \\ 
  Medium, PCA, CPs & 1.331 (12) & 1.129 (12) & 0.900 (11) & 0.132 (2) & 0.145 (1) & 0.162 (7) \\ 
   & -0.247 (12) & -0.144 (11) & -0.077 (7) & 1.256 (3) & 1.070 (2) & 1.027 (2) \\ 
  Large, PCA, CPs & 1.360 (10) & 1.138 (10) & 0.853 (12) & 0.133 (5) & 0.145 (2) & 0.163 (11) \\ 
   & -0.228 (10) & -0.111 (12) & -0.062 (12) & 1.236 (7) & 1.042 (5) & 1.027 (7) \\ 
  Small, TVPs & 1.437 (2) & 1.244 (2) & 0.971 (4) & 0.140 (12) & 0.152 (12) & 0.162 (9) \\ 
   & -0.288 (1) & -0.154 (1) & -0.040 (3) & 1.241 (12) & 1.026 (12) & 1.006 (9) \\ 
  Medium, TVPs & 1.417 (5) & 1.239 (3) & 1.007 (2) & 0.139 (11) & 0.151 (10) & 0.160 (4) \\ 
   & -0.298 (3) & -0.137 (2) & -0.031 (2) & 1.260 (9) & 1.034 (10) & 1.011 (8) \\ 
  Large, TVPs & 1.452 (1) & 1.270 (1) & 1.017 (1) & 0.134 (6) & 0.150 (9) & 0.159 (2) \\ 
   & -0.250 (5) & -0.103 (3) & -0.058 (1) & 1.204 (11) & 1.022 (11) & 1.007 (4) \\ 
  Small, PCA, TVPs & 1.364 (9) & 1.178 (8) & 0.928 (8) & 0.136 (9) & 0.149 (6) & 0.164 (12) \\ 
   & -0.274 (8) & -0.145 (7) & -0.011 (10) & 1.267 (6) & 1.049 (7) & 0.963 (12) \\ 
  Medium, PCA, TVPs & 1.368 (8) & 1.173 (9) & 0.944 (7) & 0.137 (10) & 0.150 (7) & 0.162 (10) \\ 
   & -0.262 (9) & -0.130 (9) &  0.006 (11) & 1.282 (5) & 1.056 (6) & 0.956 (11) \\ 
  Large, PCA, TVPs & 1.421 (4) & 1.230 (4) & 0.976 (3) & 0.132 (1) & 0.148 (4) & 0.161 (5) \\ 
   & -0.235 (7) & -0.094 (5) & -0.027 (4) & 1.197 (10) & 1.020 (9) & 0.998 (10) \\ 
    \midrule
\textbf{Inflation} & & & & & & \\
  Small, CPs & 0.413 (12) & 0.340 (7) & 0.271 (10) & 0.192 (4) & 0.192 (4) & 0.190 (3) \\ 
   & -0.043 (9) & -0.063 (7) & -0.114 (3) & 0.993 (2) & 0.999 (3) & 1.001 (2) \\ 
  Medium, CPs & 0.418 (10) & 0.317 (12) & 0.277 (8) & 0.192 (3) & 0.192 (3) & 0.191 (4) \\ 
   & -0.038 (8) & -0.084 (8) & -0.111 (1) & 0.989 (3) & 0.994 (4) & 0.995 (4) \\ 
  Large, CPs & 0.419 (9) & 0.320 (11) & 0.275 (9) & 0.193 (5) & 0.193 (5) & 0.191 (5) \\ 
   & -0.061 (3) & -0.072 (10) & -0.112 (2) & 1.008 (1) & 0.998 (5) & 0.995 (5) \\ 
  Small, PCA, CPs & 0.416 (11) & 0.331 (9) & 0.264 (12) & 0.190 (1) & 0.190 (1) & 0.189 (1) \\ 
   & -0.046 (7) & -0.064 (9) & -0.109 (8) & 0.963 (5) & 0.993 (1) & 0.994 (1) \\ 
  Medium, PCA, CPs & 0.424 (8) & 0.320 (10) & 0.268 (11) & 0.191 (2) & 0.191 (2) & 0.190 (2) \\ 
   & -0.010 (12) & -0.064 (11) & -0.103 (9) & 0.966 (6) & 0.997 (2) & 0.993 (3) \\ 
  Large, PCA, CPs & 0.435 (3) & 0.335 (8) & 0.310 (1) & 0.193 (6) & 0.193 (6) & 0.192 (6) \\ 
   & -0.041 (4) & -0.070 (6) & -0.070 (5) & 0.996 (4) & 0.989 (6) & 0.998 (6) \\ 
  Small, TVPs & 0.431 (5) & 0.366 (5) & 0.298 (3) & 0.204 (12) & 0.205 (8) & 0.203 (9) \\ 
   & -0.050 (2) & -0.061 (3) & -0.080 (7) & 0.997 (10) & 1.007 (9) & 0.991 (8) \\ 
  Medium, TVPs & 0.432 (4) & 0.367 (4) & 0.297 (5) & 0.204 (10) & 0.205 (10) & 0.204 (10) \\ 
   & -0.038 (5) & -0.061 (2) & -0.083 (4) & 0.999 (9) & 1.011 (8) & 0.986 (10) \\ 
  Large, TVPs & 0.428 (6) & 0.382 (2) & 0.305 (2) & 0.204 (11) & 0.205 (9) & 0.205 (12) \\ 
   & -0.061 (1) & -0.037 (4) & -0.074 (6) & 1.009 (8) & 1.006 (11) & 0.993 (9) \\ 
  Small, PCA, TVPs & 0.441 (1) & 0.379 (3) & 0.285 (6) & 0.202 (9) & 0.206 (12) & 0.204 (11) \\ 
   & -0.026 (6) & -0.051 (1) & -0.077 (10) & 0.980 (11) & 1.010 (10) & 0.978 (11) \\ 
  Medium, PCA, TVPs & 0.439 (2) & 0.384 (1) & 0.283 (7) & 0.201 (7) & 0.202 (7) & 0.200 (7) \\ 
   & -0.008 (10) & -0.028 (5) & -0.069 (11) & 0.965 (12) & 0.977 (12) & 0.950 (12) \\ 
  Large, PCA, TVPs & 0.425 (7) & 0.346 (6) & 0.298 (4) & 0.201 (8) & 0.205 (11) & 0.202 (8) \\ 
   & -0.016 (11) & -0.033 (12) & -0.046 (12) & 0.998 (7) & 1.013 (7) & 0.995 (7) \\ 
     \midrule
\textbf{Unemployment} & & & & & & \\
  Small, CPs & 1.056 (12) & 0.978 (10) & 0.818 (5) & 0.075 (12) & 0.083 (12) & 0.095 (3) \\ 
   & -0.456 (4) & -0.282 (5) & -0.041 (1) & 1.405 (10) & 1.188 (11) & 1.023 (2) \\ 
  Medium, CPs & 1.077 (11) & 1.008 (7) & 0.831 (2) & 0.074 (10) & 0.081 (10) & 0.094 (1) \\ 
   & -0.435 (5) & -0.260 (3) & -0.011 (3) & 1.388 (9) & 1.183 (9) & 1.013 (3) \\ 
  Large, CPs & 1.089 (10) & 1.028 (3) & 0.824 (3) & 0.074 (11) & 0.081 (11) & 0.094 (2) \\ 
   & -0.420 (6) & -0.234 (4) & -0.002 (4) & 1.406 (8) & 1.184 (8) & 1.022 (1) \\ 
  Small, PCA, CPs & 1.173 (2) & 1.039 (2) & 0.819 (4) & 0.073 (6) & 0.080 (9) & 0.096 (4) \\ 
   & -0.368 (1) & -0.271 (1) & -0.036 (2) & 1.415 (1) & 1.213 (3) & 1.033 (4) \\ 
  Medium, PCA, CPs & 1.139 (5) & 1.013 (6) & 0.791 (6) & 0.073 (5) & 0.080 (7) & 0.096 (5) \\ 
   & -0.400 (2) & -0.282 (2) & -0.003 (5) & 1.389 (4) & 1.193 (5) & 1.029 (5) \\ 
  Large, PCA, CPs & 1.111 (9) & 1.021 (4) & 0.840 (1) & 0.074 (8) & 0.080 (6) & 0.097 (6) \\ 
   & -0.402 (3) & -0.234 (6) &  0.077 (6) & 1.212 (12) & 1.113 (12) & 1.034 (6) \\ 
  Small, TVPs & 1.123 (8) & 0.924 (12) & 0.512 (12) & 0.071 (1) & 0.079 (2) & 0.104 (9) \\ 
   & -0.329 (12) & -0.258 (11) &  0.136 (12) & 1.356 (6) & 1.206 (1) & 1.050 (8) \\ 
  Medium, TVPs & 1.141 (4) & 0.993 (9) & 0.525 (11) & 0.072 (4) & 0.080 (4) & 0.104 (12) \\ 
   & -0.343 (10) & -0.214 (10) &  0.084 (11) & 1.393 (2) & 1.210 (2) & 1.037 (10) \\ 
  Large, TVPs & 1.127 (7) & 1.015 (5) & 0.635 (10) & 0.074 (9) & 0.080 (8) & 0.104 (11) \\ 
   & -0.343 (11) & -0.165 (12) &  0.168 (10) & 1.379 (11) & 1.195 (6) & 1.033 (9) \\ 
  Small, PCA, TVPs & 1.134 (6) & 0.956 (11) & 0.664 (9) & 0.072 (2) & 0.080 (5) & 0.104 (10) \\ 
   & -0.372 (7) & -0.254 (8) &  0.157 (8) & 1.378 (5) & 1.156 (10) & 0.964 (12) \\ 
  Medium, PCA, TVPs & 1.163 (3) & 0.994 (8) & 0.732 (8) & 0.072 (3) & 0.079 (3) & 0.101 (8) \\ 
   & -0.335 (8) & -0.214 (9) &  0.233 (9) & 1.392 (3) & 1.164 (7) & 0.981 (11) \\ 
  Large, PCA, TVPs & 1.175 (1) & 1.056 (1) & 0.758 (7) & 0.073 (7) & 0.078 (1) & 0.098 (7) \\ 
   & -0.316 (9) & -0.179 (7) &  0.175 (7) & 1.388 (7) & 1.173 (4) & 1.014 (7) \\ 
   \bottomrule
\end{tabular*}
\begin{tablenotes}[para,flushleft]
\tiny{\textit{Notes}: The first row per model shows average LPSs and RMSEs for the real time information set with the rank at the end of the holdout in parentheses. The second row indicates the difference (LPSs) and ratio (RMSEs) between real time and pseudo out-of-sample simulations within each model specification, with ranks for pseudo out-of-sample exercises in parentheses.}
\end{tablenotes}
\end{threeparttable}
\end{center}
\end{table*}

\end{appendices}
\end{document}